\documentclass[%
amsmath,amssymb,
aip,
pop%,
]{revtex4-2} %% {revtex4-1}'  

\usepackage{graphicx}% Include figure files
\usepackage{dcolumn}% Align table columns on decimal point
\usepackage{bm}% bold math
\usepackage{CJK}
\usepackage{color}

\begin{document}
%  \draft % marks overfull lines with a black rule on the right	
%  \preprint{APS/123-QED}

%\setCJKfamilyfont{Song}{SimSun}
\begin{CJK*}{UTF8}{gbsn}

%\title{Manuscript Title:\\with Forced Linebreak}% Force line breaks with \\

\title{Suppression and excitation condition of collision on instabilities of electrostatic plasmas}

%author‘s informations
%\author{Author's name}
%\affiliation{Authors' institution and/or address}
%\altaffiliation[Also at ]{Physics Department, XYZ University.}
%\email{EmailName@institution.edu}
%\thanks{}
%\homepage[]{http://www.Second.institution.edu/~Charlie.Author}

\author{Y. W. Hou(侯雅巍)}%\email{arvayhou@ahjzu.edu.cn}
\affiliation{School of Mathematics $\&$ Physics, Key Laboratory of Architectural Acoustic Environment of Anhui Higher Education Institutes, Key Laboratory of Advanced Electronic Materials and Devices,
Anhui Jianzhu University, Hefei, Anhui 230601, China}
\affiliation{CAS Key Laboratory of Geospace Environment and
Department of Plasma Physics $\&$ Fusion Engineering,
University of Science and Technology of China, Hefei, Anhui 230026, China}

\author{M. Y. Yu(郁明阳)}\email{yumingyang@sztu.edu.cn}
\affiliation{Shenzhen Technology University, Shenzhen, Guangdong 518118, China}
\affiliation{Institute for Theoretical Physics I, Ruhr University, D-44780 Bochum, Germany}

\author{J. F. Wang(王进芳)}\email{jfwang@ipp.ac.cn}
\affiliation{Institute of Plasma Physics, Chinese Academy of
Sciences, Hefei, Anhui 230031, China}

\author{C. Y. Liu(刘成岳)}%\email{cyliu@hfut.edu.cn}
\affiliation{School of Physics,
Hefei University of Technology, Hefei, Anhui 230009, China}

\author{M. X. Chen(陈美霞)}%\email{mxchen@hfut.edu.cn}
\affiliation{School of Physics,
Hefei University of Technology, Hefei, Anhui 230009, China}

\author{B. Wu(吴斌)}%\email{wubin@ipp.ac.cn}
\affiliation{Institute of Plasma Physics, Chinese Academy of
Sciences, Hefei, Anhui 230031, China}

\date{\today}

\begin{abstract}
Two-stream (TS) and Bump-On-Tail (BOT) electron distributions can induce instabilities in collisionless plasmas, which is closely related to phenomena in space and fusion plasmas. 
Collisions can lead to unexpected plasma behavior, especially in dense and/or low temperature plasmas.
In this work, the Vlasov-Poisson system with Krook collisions are used to study the effect of collisions.
Normally, the collision can dissipate the system energy which causes the suppression of the instabilities. 
Against the traditional suppression effect of collision on the instability, 
it is found in our simulation that the collision can also excite the instability even in the forbidden beam velocity range predicted by the cold-beam theory.
With collision, the beam velocity range can be divided into suppression area $[v_{th}/2, v_c+v_{th}]$, transition area $[v_c-v_{th}, v_c+v_{th}]$, excitation area $[v_c+v_{th}, 2v_c]$ and forbidden area $[2v_c, +\infty]$ for TS instability. 
where $v_c$ is the critical velocity from cold-beam theory and $v_{th}$ is thermal velocity or the beam width in our simulation. The collision dissipation effect and the excitation of beam instability can compete with each other, which evoked the excitation of collision on TS instability. The collision can change the suppression and excitation condition from beam theory. However, for BOT instability, there is only suppression effect of collision on the instability. These results can expand the view of collision effect on instability of electrostatic plasmas.
\end{abstract}

%\pacs{} PACS numbers

\maketitle %must follow title, authors, abstract, \pacs, and \keywords
%\clearpage
\end{CJK*}


\section{Introduction}
Beam-plasma instabilities can be excited when people inject beams into the plasmas \cite{Deneef1973,Whelan1983,Hartmann1995}, 
including the so called Bump-On-Tail (BOT) instability. 
When two-stream with opposite direction injected and penetrated into each other, a two-stream (TS) instability can also occur. 
Besides the Nyquist criterion\cite{Penrose1960,Hasegawa1968,Clemmow1969} for the linear instability condition, there are lots 
of analytical \cite{Infeld1970,Schamel1982,Chen1984,El-Labany1986,Rostomian1988,Boyd2003,Ng2004} and numerical \cite{Roberts1967,Morse1969b,Gentle1973a,Lacina1976,Morey1989,Zheng2006,Dieckmann2006,Daldorff2011,Volokitin2012} study to reveal the nonlinear feature of the instability. 
Recent research found that the TS and BOT instability are closely related with the electron acceleration 
and electromagnetic wave generation in space and fusion plasmas\cite{Tsiklauri2011,Thurgood2015,Thurgood2016,Alves2014,Freethy2015}.
 
Plasmas are commonly thought to be collisionless when they are rare and/or hot enough. 
However, when it comes to dense and/or low-termperature plasmas, collison can lead to behaviors which are different from collisionless cases.
On one hand, as the collision is an dissipation process, it normally can suppress beam-plasma instabilities in overdense plasmas\cite{Kemp2006} 
or increased density cases\cite{Hao2009}. On the other hand, except supression, collision can excite the instabilities under certain circumstances.
In the dense region, collision can first attenuated and then enhance the current-filamentation instability when a relativistic intense electron beam penetrates into a cold dense plasma\cite{Hao2009}. In a plasma jet, the two-stream instability (Buneman instability) can be excited by collisions based on a two-fluid collisional theory\cite{Zhou2023}. There maybe different mechanisms for the collision to suppress or excite instabilities. 
For instance, collision can induce the electromagnetic waves or plasma oscillation amplification\cite{Musha1964}. However, this collision-induced instabilities cannot occur due to run-away effect\cite{Stenflo1968} if the electron distribution becomes strong anisotropy.

From traditional point of view, the nonlinear evolution of TS or BOT instabilities can be devided into two stages, linear growth and saturation stages. 
Our simulaition results \cite{Hou2015b} show that the grow stage are not always linear due to the conbination of the wave harmonics determined by the system size and the numerical noise. 
In our past work\cite{Hou2024}, it is found that collisions can
decrease the growth rate and saturation level of the unstable waves. Collisions also
tend to shrink the phase space vortex and narrow the phase-mixed (smoothed) region
of the trapped electrons. These findings are consistent to the energy dissipation feature of collision.
However, our recent work shows that except the suppression of instability, collision can also excite the TS instability. 
To find out the suppression and excitation condition of collision on the TS instability, we take a scan of the beam velocity. 
It is found that the beam velocity range can be devided into four areas based on the feature of instability due to collision.
The transition area, during which the suppression of TS instability changes to the excitation with the beam velocity, are centered around the critical velocity  with a beam width ($[v_c-v_{th}, v_c+v_{th}]$). When the beam velocity is less than $v_c-v_{th}$, the collision can suppress the TS instability. 
When the beam velocity is in the range $[v_c+v_{th}, 2v_c]$, the collision can suppress the TS instability. When the beam velocity is bigger than $2v_c$, the TS instability can not be excited even with the collision. 

In this paper, the simulation model of one-dimensional Vlasov and Poisson system is introduced in Part II. 
In Part III, the existence conditions of the two-stream instability and the bump-on-tail instability 
from both kinetic and cold-plasma theory are given. 
The simulation results are found to be consistent with this exsistence condition.
The collision effect on the two-stream instability and the bump-on-tail instability are studied in Part IV and Part V, seperately. 

\section{The simulation model} \label{model}
The one-dimensional Vlasov and Poisson system including Krook collisions
is \cite{Chen2006}
\begin{equation}  
\partial_t f+v\partial_x f-E\partial_v f=\nu(f-f_M),   \label{e1}
\end{equation}
\begin{equation}
\partial^2_x\phi=\int^\infty_{-\infty}fdv-1,   \label{e2}
\end{equation}
where the space $x$, time $t$, electron velocity $v$, electron distribution function $f(x,v,t)$, electrostatic field $E(x,t)=-\partial_x\phi$, and potential $\phi$, have been normalized by the Debye length $\lambda_D=\sqrt{\epsilon_0k_BT/n_0e^2}$, inverse plasma frequency $\omega^{-1}_p=\sqrt{m\epsilon_0/n_0e^2}$, electron thermal speed $v_T=\sqrt{k_BT/m}$, $m\lambda_D\omega_p^2/e$, and $m\lambda_D^2\omega_p^2/e$, respectively. Here, $-e$, $m$, $T$, $n_0$, $\epsilon_0$, $k_B$, $\nu$, and $f_M(v)=(2\pi)^{-1/2}\exp(-v^2/2)$ are the electron charge, mass, temperature, background plasma density, the vacuum permittivity, Boltzmann constant, collision rate, and the Maxwell-Boltzmann distribution, respectively. The collision time is thus $1/\nu$. Overall charge neutrality of the plasma is maintained by a uniform stationary ion background.

For investigating the two-stream and bump-on-tail instabilities, we invoke the bi-Maxwellian background electron distribution
\begin{equation}  %\end{equation}
f_0(v)=C(2\pi)^{-1/2}\exp[-(v-v_1)^2/2]+(1-C)(2\pi)^{-1/2}\exp[-(v+v_2)^2/2)], \label{e3}
\end{equation}
where $C$ and $1-C$ are the relative concentrations, and $v_1$ and $-v_2$ the center speeds, of the two electron groups.

Periodic boundary conditions $f(x,v,t)=f(x+L,v,t)$ and $\phi(x,t)=\phi(x+L,t)$ are used, where $L$ is the length of simulated region. Thus, phenomena with characteristic lengths larger than $L$ are precluded. To avoid relativistic effects, we set $f(x,v,t)=0$ for $|v|>v_m$, where $v_m=8$ for considering Landau damping and $v_m=10$ for considering the instabilities. (That is, $v_m\ll c$ and $c$ is the light speed normalized by $v_T$.) A fourth-order Runge-Kutta scheme is used to solve the Vlasov equation in time and the centered-difference method is used to advance $x$ and $v$ \cite{Cheng1976,Hou2011a,Hou2011b}. The Thomas algorithm together with the Simpson method for the integration is used in solving the discretized Poisson equation with the centered-difference method. Without the collision term, the simulation code is fourth order accurate in time and second order accurate in space and velocity. The simulation code has been bench-marked by comparing its results on linear and NLLD, as well as plasma wave echoes and instabilities \cite{Hou2011a,Hou2011b,Hou2015a,Hou2015b,Hou2016}, with that of existing numerical and analytical works.

\section{Existence condition of the two-stream instability and the bump-on-tail instability}
The linear behaviors of the instabilities can be achieved 
by eigenmode analysis based on dispersion relation.
The dispersion relation\citep{Clemmow1969,Chen1984} $D(k,\omega)=0$ describes the relation 
between wave vector $k$ (spatial feature) and frequency $\omega$ (time feature) 
for electrostatic waves in an electron plasma.
The frequency can be divived into two parts $\omega=\omega_r+i\gamma$, the real frequency $\omega_r$ and growth/damping
rate $\gamma$, which show the oscilating and growth/damping feature seperately.
The dispersion relation is closely related to the electron plasma distribution function based on kinetic theory, 
\begin{equation}
	D(k,\omega)=1-\frac{e^2}{\epsilon_0mk^2}
	\int_{-\infty}^{\infty}\frac{d_vf_0(v)}{v-\omega/k}dv.
\end{equation}
When $\gamma\ll\omega_r$, $\gamma$ can be given by an analytical expression
\begin{equation}
	\gamma=-(\pi\omega_p/2k)d_vf_0(v)|_{v=\omega_r/k}(\omega_r/k-d_k\omega_r).
\end{equation}
When $d_vf_0(v)>0$, a mode with real frequency $\omega_r$ can grow, 
which is so called inverse Landau damping. 
This mode growth can be attributed to energy exchange of resonance between the wave and the
electrons, whose velocities is close to the wave phase velocity $v\sim\omega_r/k$.

\subsection{Existence condition of the two-stream instability}
To simplify the estimation, the cold plasma limit, which replace the
gaussian functions in $f_0$ by delta functions, can be used to 
estimate the existence condition and growth rate of the beam instability
when thermal effects do not contribute significantly to the instability
\citep{Clemmow1969}. 
The dispersion relation of the cold plasma limit for two-streams is
\citep{Clemmow1969,Chen1984,Boyd2003}
\begin{equation}
	\frac{\omega_{pa}^2}{(\omega-kv_{b})^2}+\frac{\omega_{pb}^2}{(\omega+kv_{b})^2}=1,
\end{equation}
where $\omega_{pa}=\sqrt{C}\omega_p$ and
$\omega_{pb}=\sqrt{1-C}\omega_p$ are the plasma frequency of two-streams.

The existence condition of beam instability is $0<k<k_{\rm c}$, where
\begin{equation}
	k_{\rm cr}=[(C^{1/3}+(1-C)^{1/3})^{3/2}]\omega_p/2v_b.
\end{equation}

The wavevector/wavelength of the noise-excited mode in the simulations 
is located among $2\pi/L<k<2M\pi/L$ or $L/M<\lambda<L$, 
where $L$ is the system length and $M$ is the grid number of the system. 
In this study, $M$ is set to be $2000$, so $2\pi/L \ll 2M\pi/L$ or $L/M \gg L$. 
The existence condition for instability can be expressed by wave vector, beam velocity or system length. 
$k_{\rm cr}\geq k_{\min}$, where $k_{\min}=2\pi/L$ is the minimum wave vector
allowed in the simulation, that is, $v_b\leq v_{c}$, where the
critical velocity is
$v_{c}=(C^{1/3}+(1-C)^{1/3})^{3/2}\omega_pL/4\pi$, 
or $L\geq L_{c}$, where the critical system length is
$L_{c}=4\pi(C^{1/3}+(1-C)^{1/3})^{-3/2}v_b/\omega_p$.

\begin{figure}%[htbp]%[p!]%
\includegraphics[height=5cm,width=6cm]{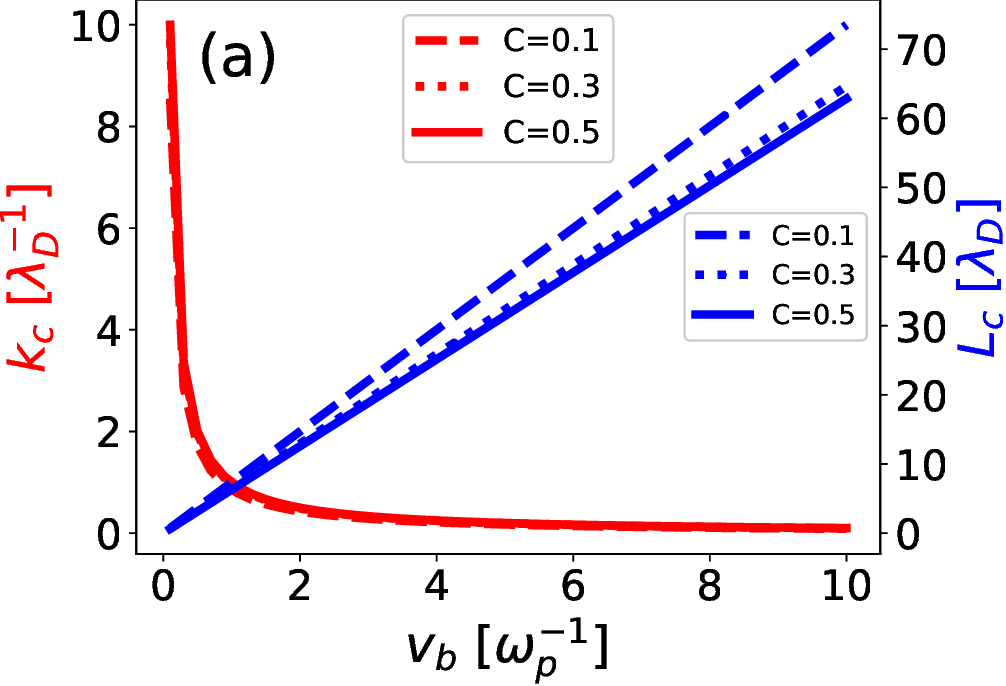}
\includegraphics[height=5cm,width=6cm]{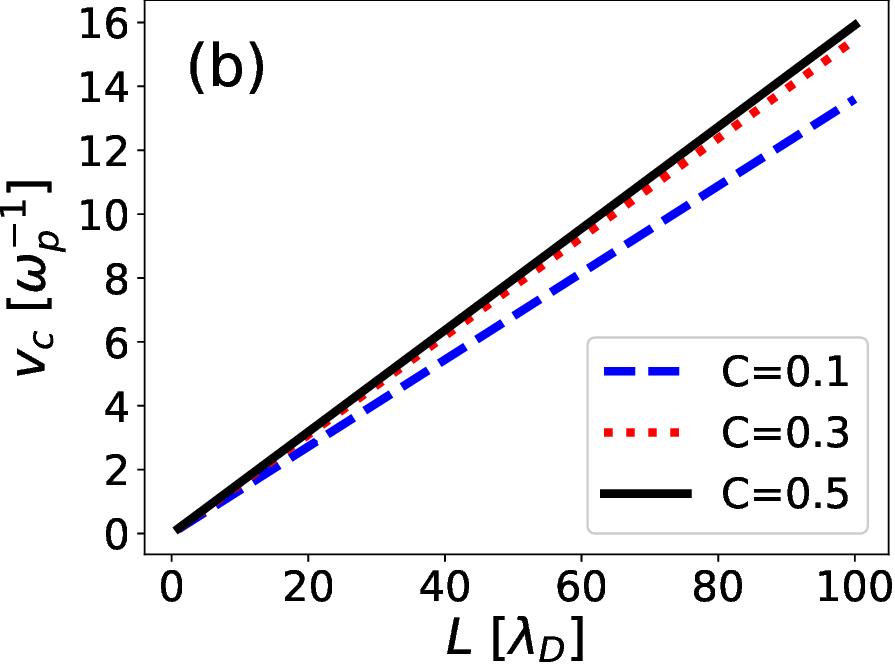}
\caption{The critical threshold of the two-stream instability based on cold-beam theory. 
		 (a) Critical wave vector $k_c$ and critical system length $L_c$ dependent on beam velocity $v_b$.
		 (b) Critical beam velocity $v_0$ dependent on system length $L$.
\label{Fig1}}
\end{figure}

The critical wave vector $k_c$ and critical system length $L_c$ dependent on beam velocity $v_b$ are plotted out in Fig.\ \ref{Fig1}(a) 
for different beam strength $C$. It is found that with the increase of beam velocity, the critical vector decreases rapidly 
and critical system length increases linealy. The beam strength doesn't inflence the critical vector significantly. 
The influence of beam strength on the criticl system length increases with the beam velocity. 
From Fig.\ \ref{Fig1}(b), one can find that critical velocity increases with the system length. 
The influence of beam strength on the criticl velocity also increases with the system length. 

\begin{figure}%[htbp]%[p!]%
\includegraphics[width=4.5cm]{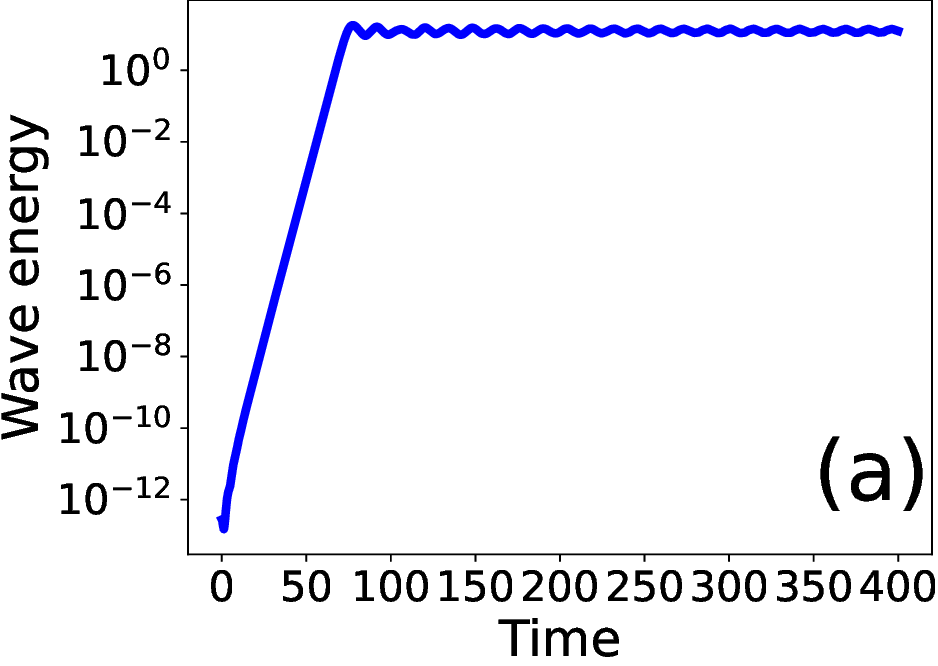}\includegraphics[width=4.5cm]{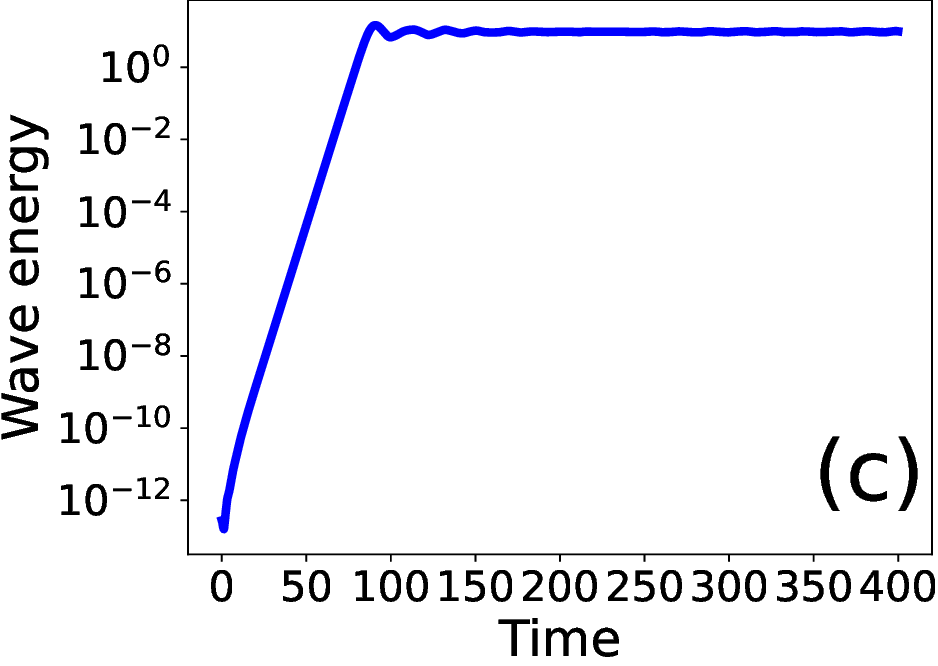}\includegraphics[width=4.5cm]{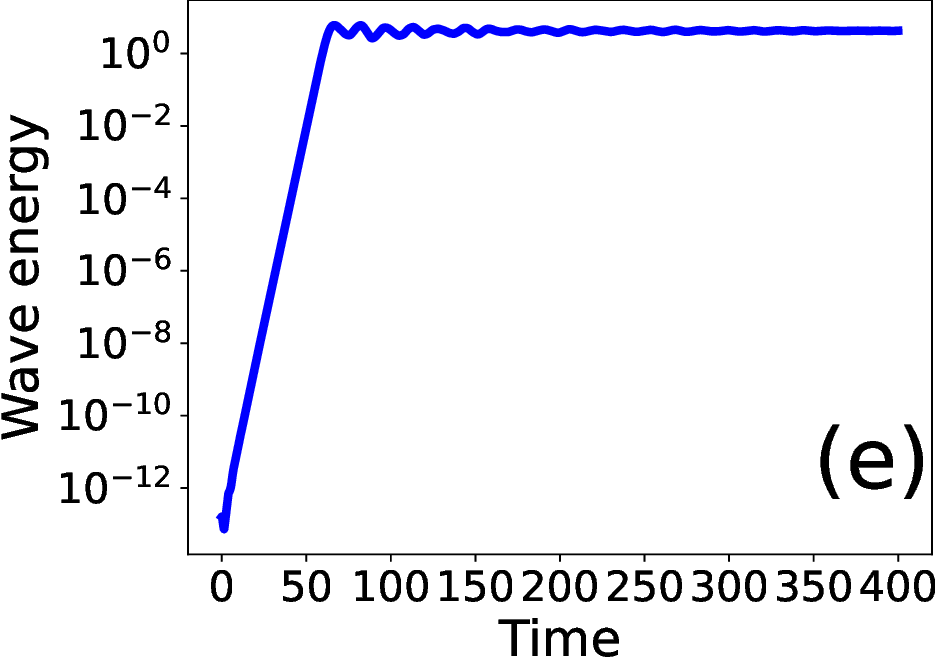}	
\includegraphics[width=4.5cm]{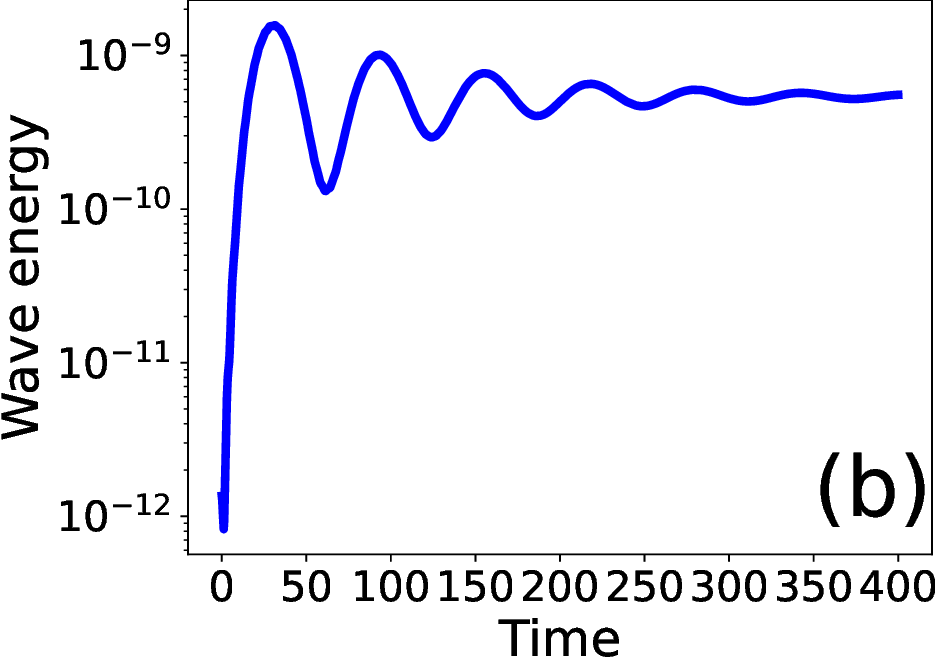}\includegraphics[width=4.5cm]{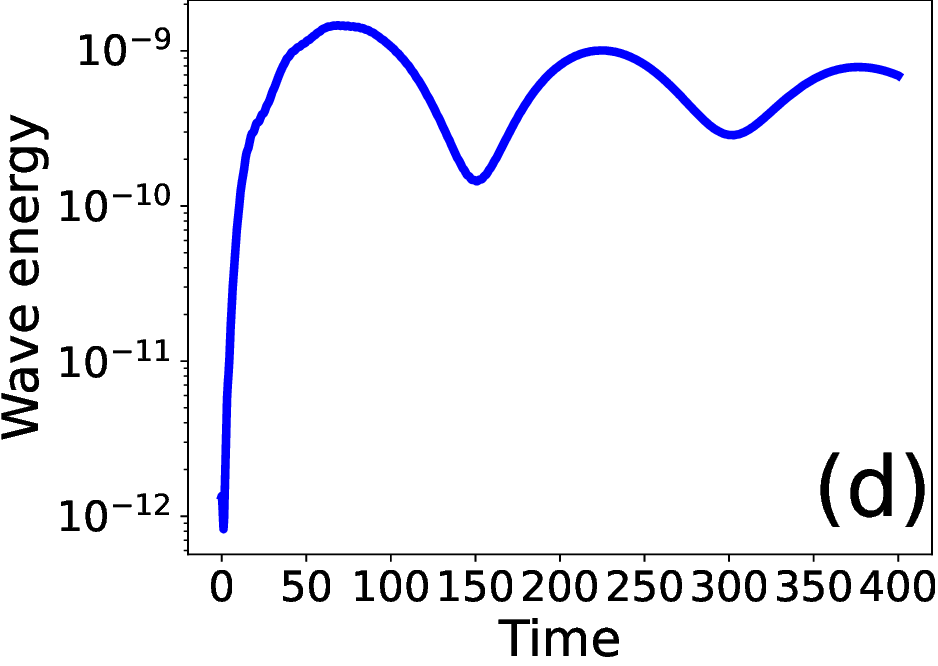}\includegraphics[width=4.5cm]{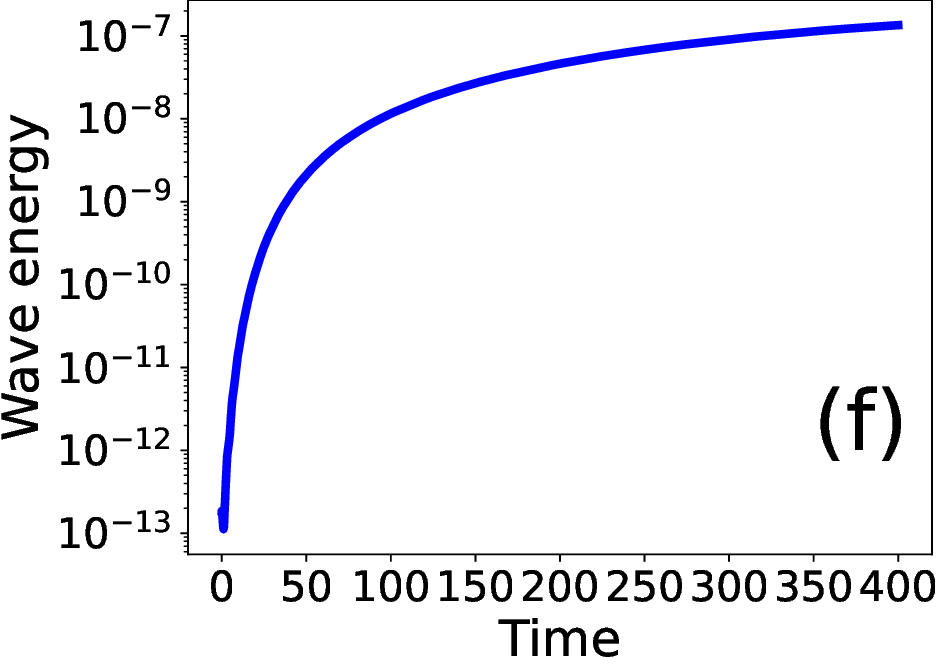}		
\caption{Evolution of the wave energy for (a) $L=8\pi$, $C=0.5$ and $v_b=4$,
		,(b) $L=8\pi$, $C=0.5$ and $v_b=4.5$, (c) $L=8\pi$, $C=0.3$ and $v_b=4$, (d) $L=8\pi$, $C=0.3$ and $v_b=4.5$,
		(e) $L=7\pi$, $C=0.5$ and $v_b=3$,(f) $L=7\pi$, $C=0.5$ and $v_b=4$.
\label{Fig2}}
\end{figure}

To check the applicability of the instability exsistence condition from the cold-plasma theory, 
especially the linear growth behavior of the instabilities, we perform the simulations 
of the TS instability with different parameters.
For $L=8\pi$, $C=0.5$ and $v_b=4$, the critical velocity is $v_c=4.0$ 
and the time evolution of wave energy is shown in Fig.\ \ref{Fig2}(a). 
One can see that the wave energy grow exponentially first and then oscillate after saturation.
If we change the beam velocity to $v_b=4.5$ which is bigger than the critical velocity, 
the wave energy (Fig.\ \ref{Fig2}(b)) grow first and then saturate at an very low level $10^{-9}$, 
which is $9$ orders of magnitude lower than the first case. 
The critical velocity is deduced from cold beam theory and the width of beam is ignored.
In our simulation, the velocity is normalized by the thermal velocity of electron, which is the width of beam in our initial setting, 
so the applicable critical velocity is the critical velocity plus $0.5$ or $1$.
As can be seen in Fig.\ \ref{Fig2}(b), it is hard to excite two-stream instability which is consistent with the results from the theory.
To see the effect of beam strength on critical velocity, we change $C$ to $0.3$, then the corresponding critical velocity should be $v_c=3.9$, 
which is a little smaller than the initial case. It is found that the behavior of the wave energy evolution is similar to that with $v_c=4.0$ 
for $v_b=4.0$ (Fig.\ \ref{Fig2}(c)) and $v_b=4.5$ (Fig.\ \ref{Fig2}(d)).
The system length can also affect the critical velocity, so we change system length to $L=7\pi$ and the critical velocity will be $v_c=3.5$.
For beam velocity $v_b=3.0$, which is less than the $v_c$, the TS instability can be excited (Fig.\ \ref{Fig2}(d)) with the exponential growth linear stage and oscilating saturation stage. However, when the beam velocity is set to be $v_b=4.0$(Fig.\ \ref{Fig2}(d)), the wave energy grows slowly and approach to a low level $10^{-7}$. These results show that system length, beam velocity and beam strength can affect the excitation of TS instability. 
The criterion of stability condition from cold-beam dispersion relation can be applicable to the simulation when the beam width can be ignored or the beam width is very small compared with beam velocity. If the beam velocity is not too large compared to the beam width, the critical velocity in use should include the beam width effect.

The parameters related to the criterion of the instability and the simulation results are shown in Table \ref{Table1}.
When the critial velocity is $4.0$ with system length $L=8\pi$ and beam strength $C=0.5$, 
the TS instability can be excited with beam velocity $v_b=4.0$, 
but it is hard to be excited with $v_b=4.5$. If we change beam strength to be $C=0.3$, then the critical velocity would be $v_c=3.9$. 
It is also hard to excite TS instability with $v_b=4.5$ and easy to excite TS instability with $v_c=4.0$.
If we change to system length to be $L=7\pi$, the velocity would be $v_c=3.5$.
It is found that TS instability is hard to be excited with $v_b=4.5$ and easy to be excited with $v_c=4.0$.
From these results, we can get to the conclusion that if the beam velocity is large than the critical velocity and a half width of the beam, 
there is a big chance that the instability is hard to be excited. 

\begin{table*}
	\caption{Parameters set for two-stream instability.
		System length $L$, beam strength $C$, critical velocity $v_c$ 
		and beam velocity $v_b$ are the four parameters in 
		the criterion of the instability.}
    \label{Table1}
\begin{ruledtabular}
	\begin{tabular}{ccccccc}
		$L$     &  $C$  &  $v_c$  & $v_b$ &  $instability$ & $Fig.\ \ 2$ \\  \hline
		$8\pi$  &  0.5  &  4.0    &  4.0  &  $\checkmark$  & $(a)$    \\
		$8\pi$  &  0.5  &  4.0    &  4.5  &  $\times$      & $(b)$    \\
		$8\pi$  &  0.3  &  3.9    &  4.0  &  $\checkmark$  & $(c)$    \\
		$8\pi$  &  0.3  &  3.9    &  4.5  &  $\times$      & $(d)$    \\
		$7\pi$  &  0.5  &  3.5    &  3.5  &  $\checkmark$  & $(e)$    \\
		$7\pi$  &  0.5  &  3.5    &  4.0  &  $\times$      & $(f)$    \\
	\end{tabular}
\end{ruledtabular}
\end{table*}

\subsection{Existence condition of the bump-on-tail instability} 
Similarly, the linear dispersion relation for Bump-On-Tail instability from cold-beam theory is
\begin{equation}
	\frac{\omega_{pa}^2}{(\omega-kv_{b})^2}+\frac{\omega_{pb}^2}{\omega^2}=1,
\end{equation}
where $\omega_{pa}=\sqrt{C}\omega_p$ and
$\omega_{pb}=\sqrt{1-C}\omega_p$ are the plasma frequency of the beam and background plasma.

The BOT instability appears for $0<k<k_{\rm c}$, where
\begin{equation}
	k_{\rm cr}=[(C^{1/3}+(1-C)^{1/3})^{3/2}]\omega_p/v_b.
\end{equation}

In the simulations, the existence condition for the instability is then 
$k_{\rm c}\geq k_{\min}$, where $k_{\min}=2\pi/L$ is the minimum wave vector
allowed in the simulation, that is, $v_b\leq v_{c}$, where the
critical velocity is
$v_{c}=(C^{1/3}+(1-C)^{1/3})^{3/2}\omega_pL/4\pi$, 
or $L\geq L_{cr}$, where the critical system size is
$L_{c}=4\pi(C^{1/3}+(1-C)^{1/3})^{-3/2}v_b/\omega_p$.

\begin{figure}%[htbp]%[p!]%
\includegraphics[height=5cm,width=6cm]{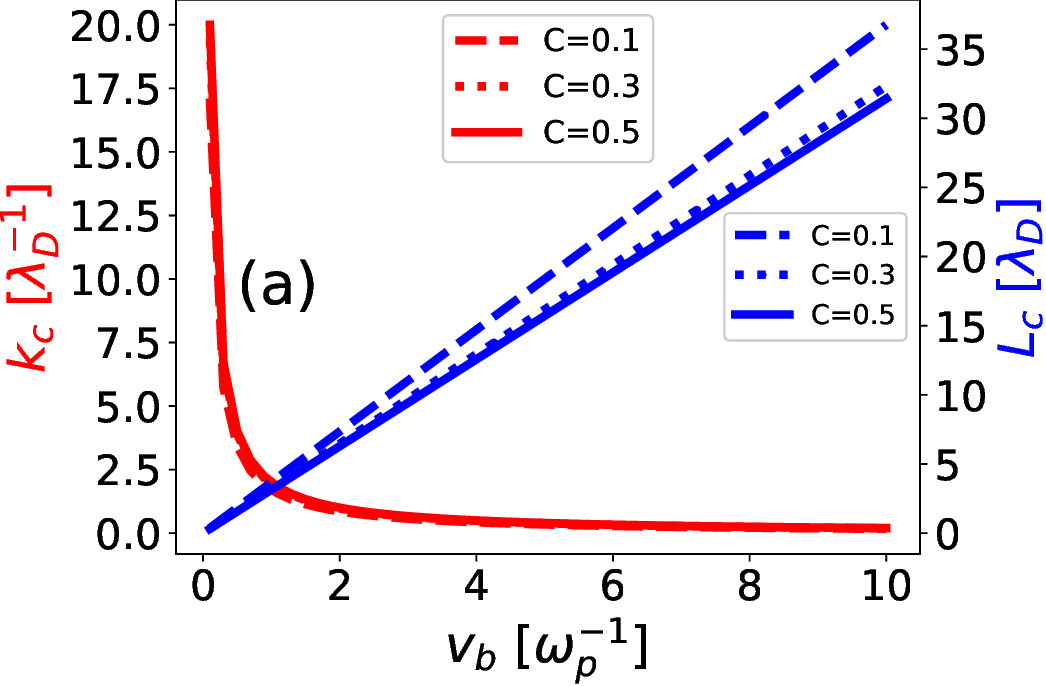}
\includegraphics[height=5cm,width=6cm]{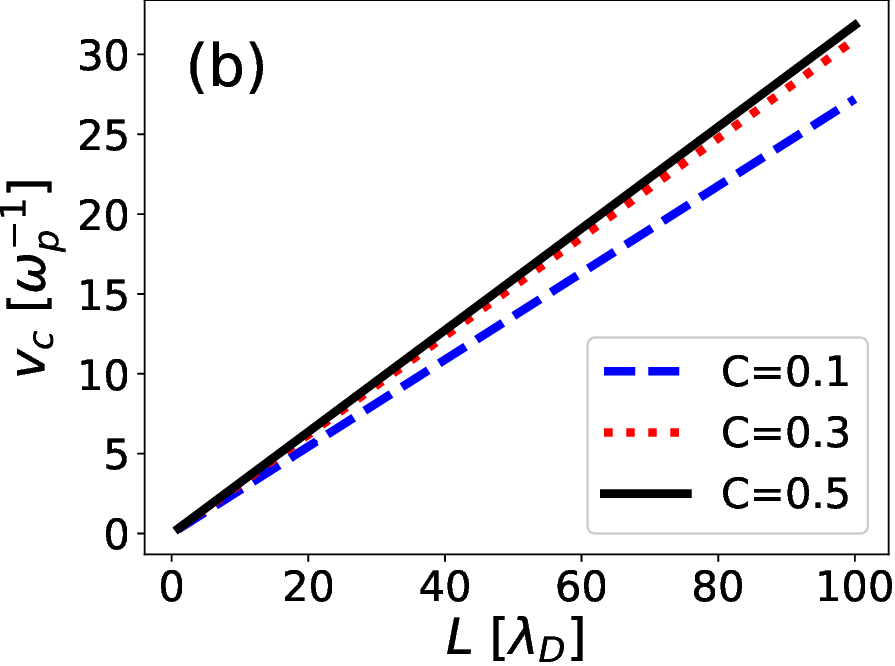}
\caption{The critical threshold of the Bump-On-Tail instability based on cold-beam theory. 
		(a) Critical wave vector $k_c$ and critical system length $L_c$ dependent on beam velocity $v_0$.
		(b) Critical beam velocity $v_0$ dependent on system length $L$.
\label{Fig3}}
\end{figure}

\begin{figure}%[htbp]%[p!]%
\includegraphics[width=4.5cm]{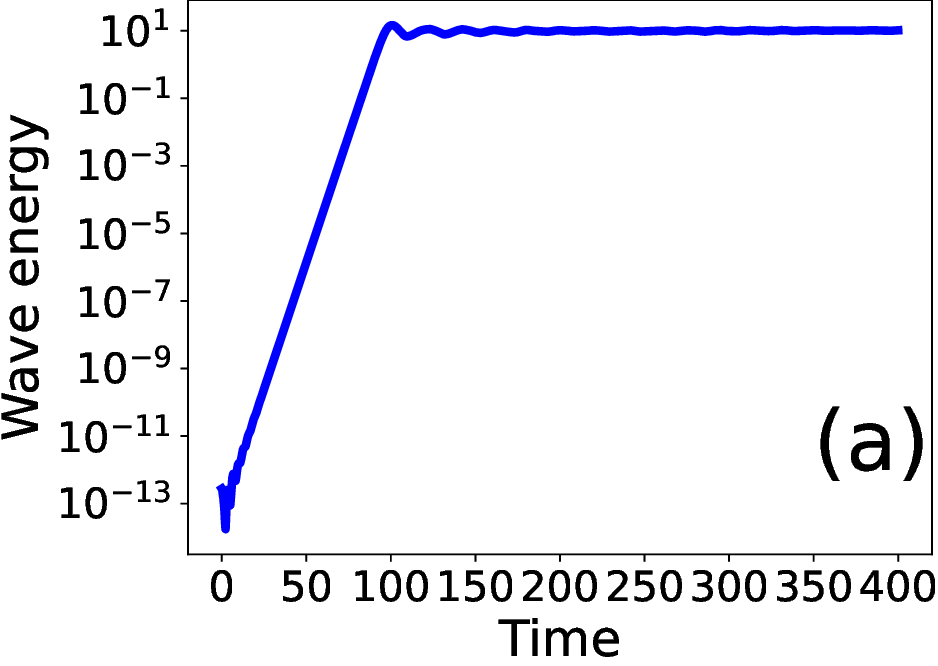}\includegraphics[width=4.5cm]{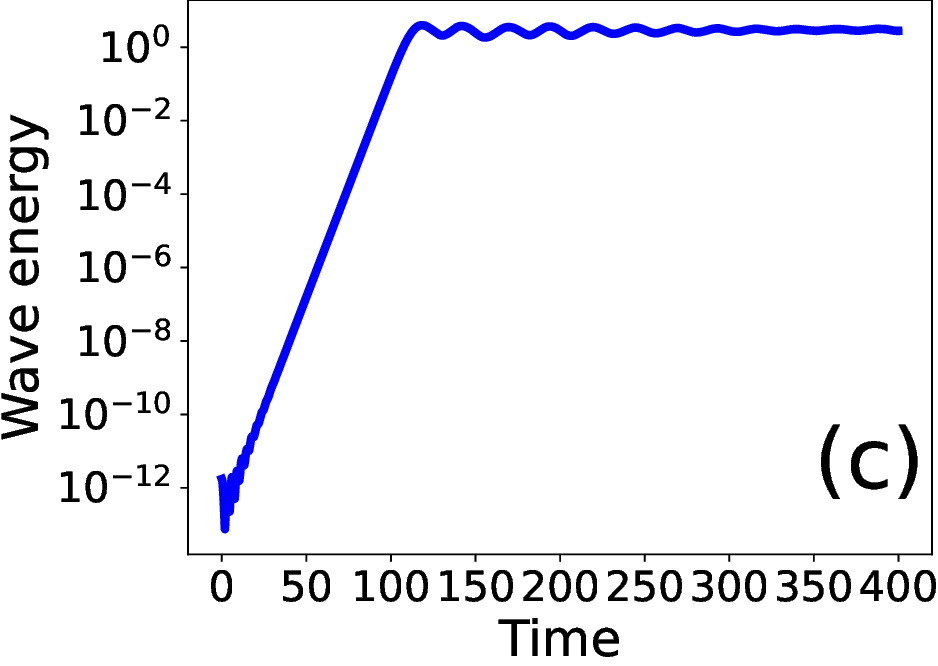}\includegraphics[width=4.5cm]{fig2e.eps}
\includegraphics[width=4.5cm]{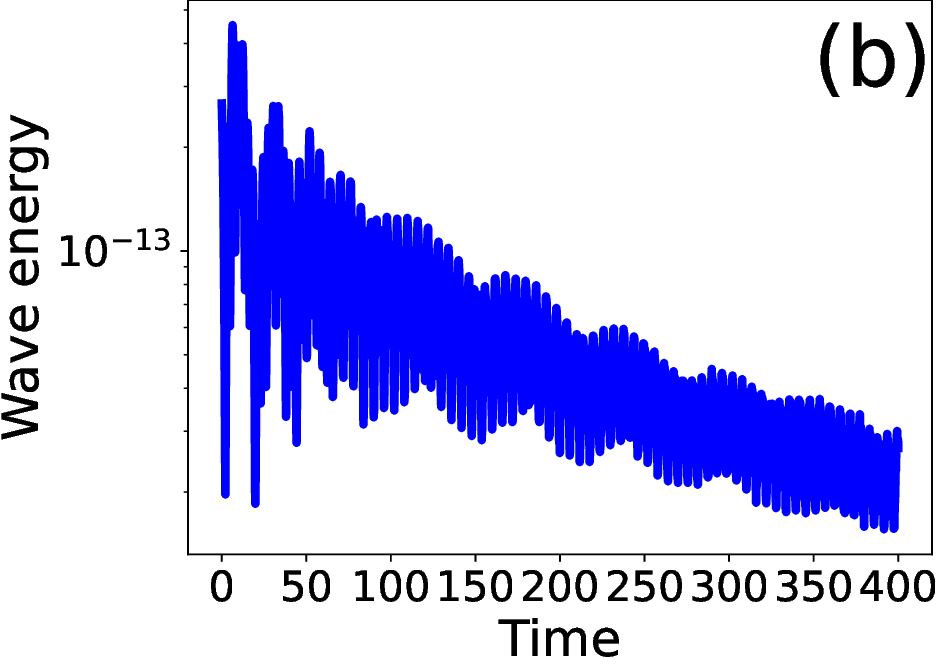}\includegraphics[width=4.5cm]{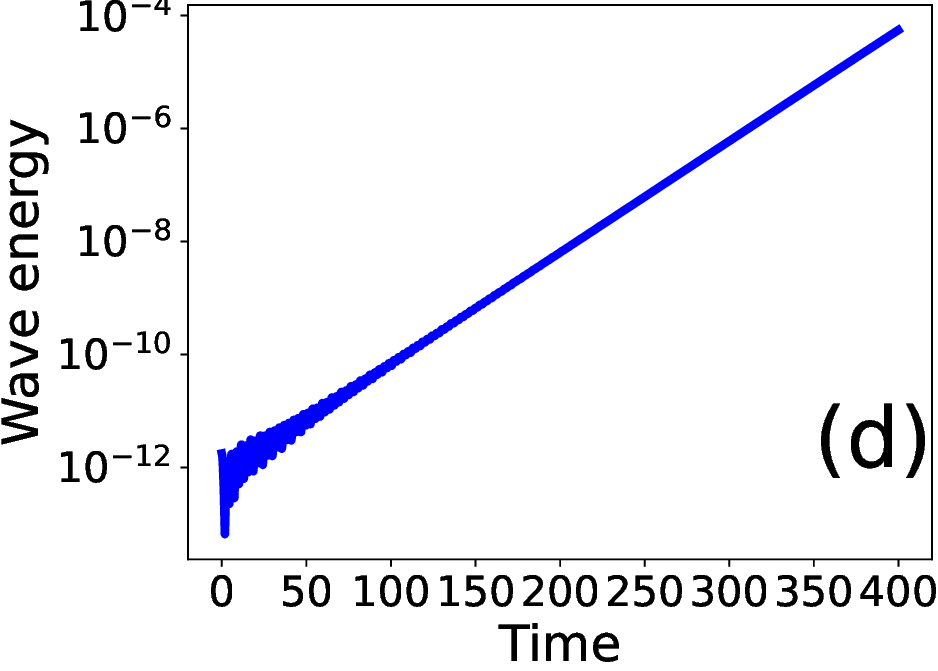}\includegraphics[width=4.5cm]{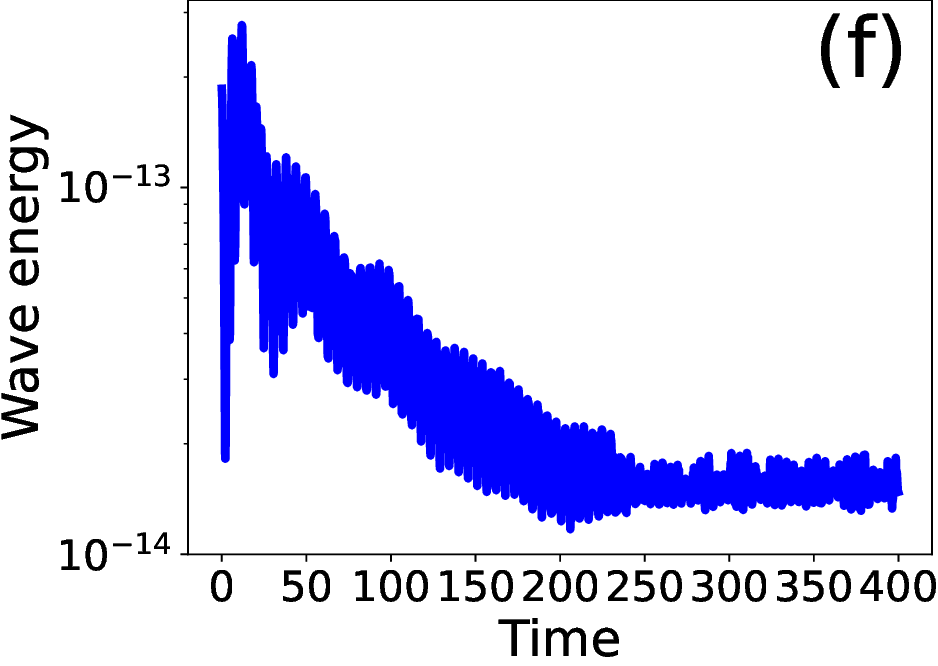}
\caption{Evolution of the wave energy for (a) $L=8\pi$, $C=0.7$ and $v_b=8$,
	,(b) $L=8\pi$, $C=0.7$ and $v_b=95$, (c) $L=8\pi$, $C=0.9$ and $v_b=7$, (d) $L=8\pi$, $C=0.9$ and $v_b=8$,
	(e) $L=7\pi$, $C=0.7$ and $v_b=7$,(f) $L=7\pi$, $C=0.7$ and $v_b=8$. 
\label{Fig4}}
\end{figure}

We also change the parameters of the BOT instability simulations 
to check the applicability of the instability exsistence condition based on cold-beam limit.
For $L=8\pi$, $C=0.5$ and $v_b=8$, the critical velocity is $v_c=7.8$ 
and the time evolution of wave energy is shown in Fig.\ \ref{Fig4} (a). 
One can see that the wave energy grow exponentially first and then oscillate after saturation.
If we change the beam velocity to $v_b=9.0$ which is bigger than the critical velocity plus beam velocity width, 
the wave energy (Fig.\ \ref{Fig4}(b)) oscillate at a very low level $10^{-13}$, 
which means no instability is excited. 
The results in Fig.\ \ref{Fig4} (a) and (b) are consistent with that from the theory, because $v_b=8$ meets the instability criterion, but $v_b=9$ does not.
To see the effect of beam strength on critical velocity, we change $C$ to $0.1$, then the corresponding critical velocity should be $v_c=6.8$. 
When $v_b=7.0$, the BOT instability (Fig.\ \ref{Fig4}(c)) is excited with the initial linear growth stage and saturation stage. When $v_b=8.0$, the BOT instability (Fig.\ \ref{Fig4}(d)) grow very slow which means it's hard to be excited.
The system length can also affect the critical velocity, so we change system length to $L=7\pi$ and the critical velocity will be $v_c=6.8$.
For beam velocity $v_b=7.0$, which is little bigger than the $v_c$, but not bigger than $v_c+v_{th}$, the BOT instability can be excited (Fig.\ \ref{Fig4}(e)) with the exponential growth linear stage and oscilating saturation stage. However, when the beam velocity is set to be $v_b=8.0$(Fig.\ \ref{Fig4}(f)), the wave energy grows and decrease at a low level $10^{-13}$, which means no BOT instability is excited. These results show that system length, beam velocity and beam strength can also affect the excitation of BOT instability. 
Since the beam velocity is not too large compared to the beam velocity width, the critical velocity in use should also include the beam velocity width effect.

The parameters related to the criterion of the instability and the simulation results are shown in Table \ref{Table2}.
When the critial velocity is $7.8$ with system length $L=8\pi$ and beam strength $C=0.3$, 
the BOT instability can be excited with beam velocity $v_b=8.0$, 
but it can not be excited with $v_b=9.0$. If we change beam strength to be $C=0.3$, then the critical velocity would be $v_c=6.8$. 
It is also hard to excite TS instability with $v_b=8.0$ and easy to excite TS instability with $v_c=7.0$.
If we change to system length to be $L=7\pi$, the vilocity would be $v_c=6.8$.
It is found that BOT instability can not be excited with $v_b=8.0$ and easy to be excited with $v_c=7.0$.
For the BOT instability, if the beam velocity is large than the critical velocity and one beam velocity width, 
normally the instability is hard to be excited. 

\begin{table*}
	\caption{Parameters set for Bump-On-Tail instability.
             System length $L$, beam strength $C$, critical velocity $v_c$ 
             and beam velocity $v_b$ are the four parameters in 
             the criterion of the instability.}
    \label{Table2}
	\begin{ruledtabular}
		\begin{tabular}{ccccccc}
			$L$     &  $C$  &  $v_c$  & $v_b$ &  $instability$ & $Fig.\ \ 4$ \\  \hline
			$8\pi$  &  0.3  &  7.8    &  8.0  &  $\checkmark$  & $(a)$    \\
			$8\pi$  &  0.3  &  7.8    &  9.0  &  $\times$      & $(b)$    \\
			$8\pi$  &  0.1  &  6.8    &  7.0  &  $\checkmark$  & $(c)$    \\
			$8\pi$  &  0.1  &  6.8    &  8.0  &  $\times$      & $(d)$    \\
			$7\pi$  &  0.3  &  6.8    &  7.0  &  $\checkmark$  & $(e)$    \\
			$7\pi$  &  0.3  &  6.8    &  8.0  &  $\times$      & $(f)$    \\
		\end{tabular}
	\end{ruledtabular}
\end{table*}

\section{Supression and excitation of collision on the two-stream instability}
\begin{figure}%[htbp]%[p!]%
\includegraphics[height=5cm,width=6cm]{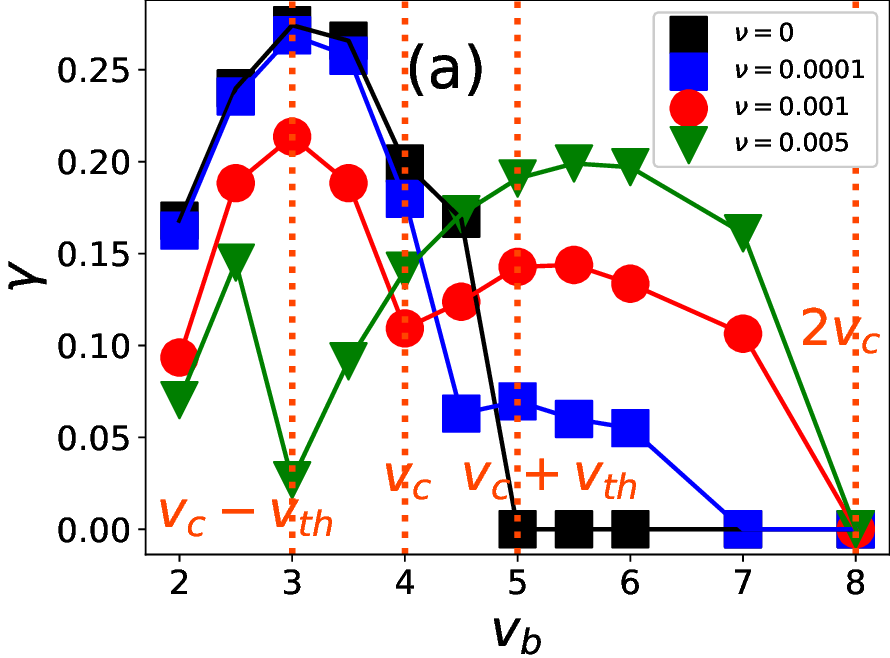}
\includegraphics[height=5cm,width=6cm]{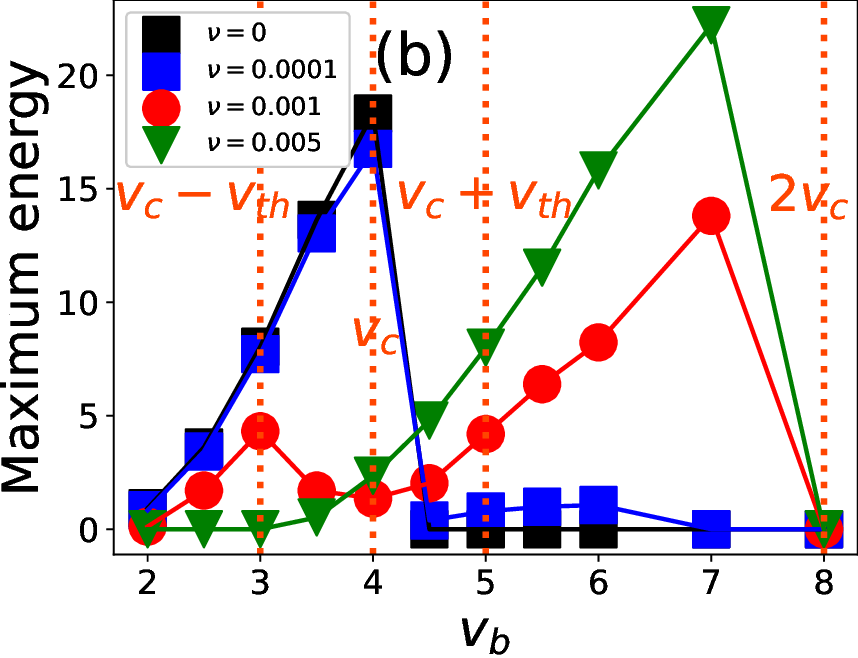}
\caption{The growth rate (a) and maximum energy (b) of the two-stream instability depend on on beam velocity $v_b$. 
\label{Fig5}}
\end{figure}

Without collision, the growth rate of TS instability increases first and then decreases with the beam velocity (Fig.\ \ref{Fig5}). 
The maximum wave energy increases with the beam velocity and directly approaches to $0$ 
for beam velocity bigger than the critical velocity plus a half beam width. 

The growth rate and maximum wave energy get to maximum value at $v_b=3$ and $v_b=4$ seperately, then approaches to $0$ for beam velocity bigger than $4.5$ and $5.0$. Since the critical beam velocity is $v_c=4.0$ from the cold-beam theory, we can see that the threshhold for the instability to occur is critical beam velocity plus one beam width.
From the Fig.\ \ref{Fig5} (a), as the beam velocity increases from $2$ to $3$, the collision decreases the growth rate. As the beam velocity increases from $5$ to $8$, the collision increases the growth rate. There is transition range for beam velocity between $3$ and $5$, which is the ciritical velocity minus and plus one beam width in phase space. During this transition area, the instability excited by the collison with $\nu=0.005$, $\nu=0.001$ and $\nu=0.0001$ from the beam velocity $v_b=3.0$, $v_b=4.0$ and $v_b=4.5$ seperately. In the collision excitation area for beam velocity between $5$ and $8$, the growth rate increases first and then decreases. For beam velocity equal or bigger than $8$, collision can not excite the TS instability.

From the Fig.\ \ref{Fig5} (b), in the suppression area for beam velocity between $2$ and $3$, the maximum wave energy decreases with the increase of collision. In the excitation area for beam velocity between $4.5$ and $8.0$, the maximum wave energy increases with the increase of collision. In the transition area for beam velocity between $3$ and $5$, after suppression, the maximum wave energy start to increase for the collison with $\nu=0.005$, $\nu=0.001$ and $\nu=0.0001$ from the beam velocity $v_b=3.0$, $v_b=4.0$ and $v_b=4.5$ seperately.

The results showed in Fig.\ \ref{Fig5} can be understood by the collision effect and the criterion of instability. 
When the beam velocity is smaller than the critical velocity, the growth rate increases and decreases, 
however the maximum energy increases with the beam velocity.  
Normally, the collision can dissipate energy from the whole system, so the growth rate and the maximum energy decreases with collision during the suppression area of beam velocity. The collision can change the plasma distribution function to Maxwellian with time. It can be considered as a small beam with beam velocity $v_b=0$ grows gradually with time. With the grow of the center beam caused by the collision, the two-stream system becomes to three-stream system. The center beam can interact with the initial beams whose beam velocity is in two opposite directions. This three-stream system can be considered as combination of two two-stream systems, so the behavior changes. For the beam velocity bigger than the critical velocity, the excitation of the instability by collision can be thought that the collision generate a center beam and change the system instability criterion. The initial beam velocity changes to be half of itself by the collision, so that the collision changes the beam velocity to meet the criterion of the instability when the initial beam velocity is between $v_c$ and $2v_c$. When the initial beam velocity is bigger than $2v_c$, even with collision, the collision changed beam velocity is still bigger than critical velocity $v_c$, which means the instability criterion can not be satisfied, so the instability can not be excited by the collision.

\begin{figure}%[htbp]%[p!]%
\includegraphics[height=5cm,width=6cm]{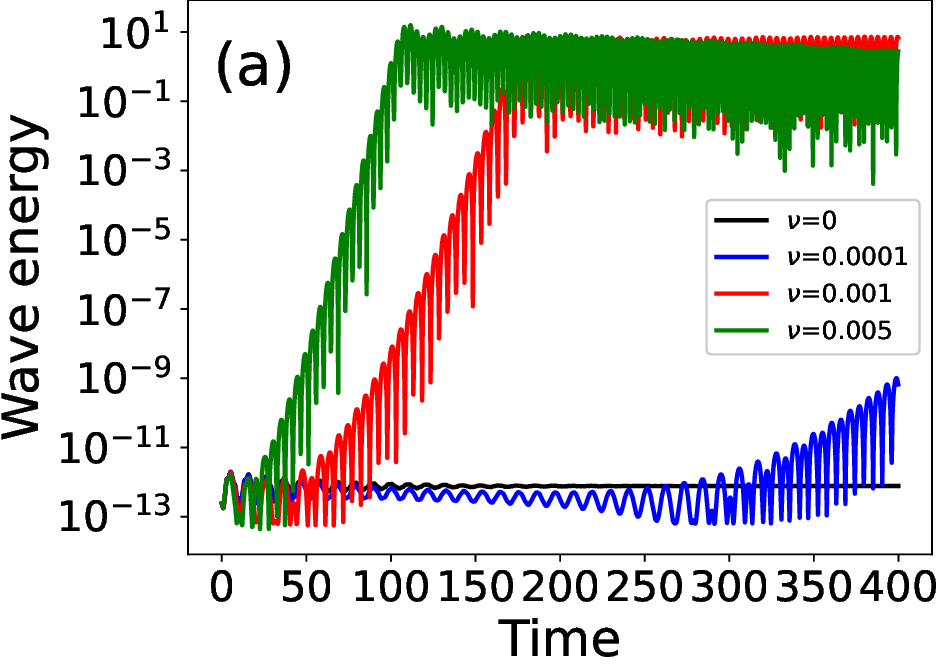}
\includegraphics[height=5cm,width=6cm]{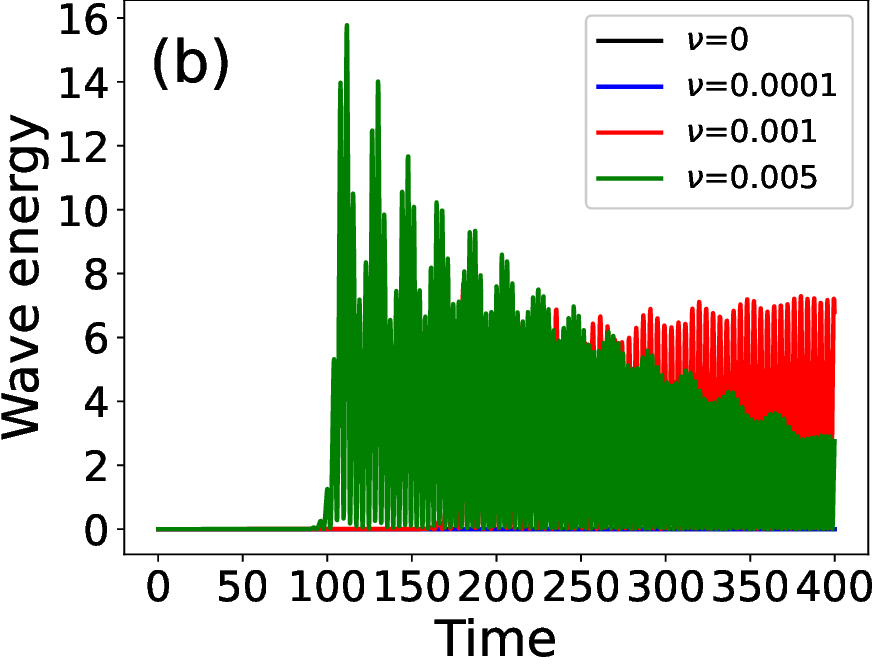}
\caption{Evolution of the excited-wave energy of the two stream instability for $v_1=-6$ and $v_2=6$, 
	     and collision rates $\nu=0$ (black), $0.0001$ (red), $0.001$ (blue) and $0.005$ (green), 
	     in (a) log and (b) normal scales.  
\label{Fig6}}
\end{figure}

To investigate the excitation effect of collision, the TS instability with beam velocity $v_b=6.0$ are carefully studied. The time evolution of wave energy with different collision rates are shown in Fig.\ \ref{Fig6}. When the collision rate is small, it is hard to excite the instability. As the increase of the collision rate, both the growth rate and maximum energy become bigger. In the linear stage, one can see that the wave energy grows exponentially with oscilation. This oscilation is caused by the interaction with center beam caused by collision and initial two-beams. The oscilation is one of the features for multi-beam instability. However after maximum energy, the wave energy with stronger collision $\nu=0.005$ decreases with time more quickly than that with weaker collision $\nu=0.001$. That is, although the strong collision can excite the beam quickly at the initial stage, it can also dissipate the wave energy quickly after the maximum energy or during the "saturation" stage.

\begin{figure}%[htbp]%[p!]%
\includegraphics[width=15cm,height=7.5cm]{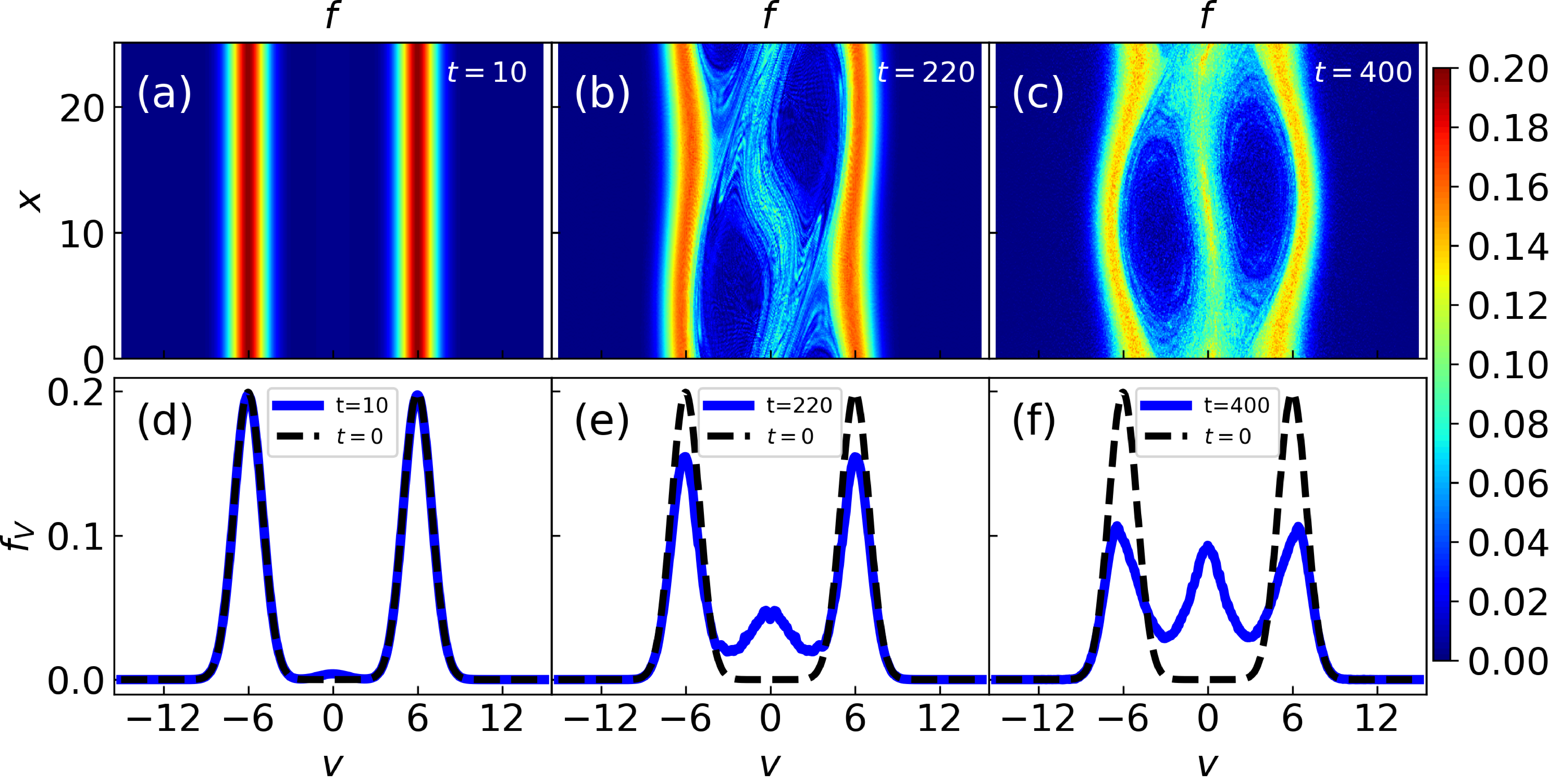}
\caption{The electron phase space (upper row) and $f_{\rm v}$ (lower row) of the two-stream instability for $\nu=0.001$ at $t=10$ [(a) and (d)], $220$ [(b) and (e)], and $400$ [(c) and (f)].
\label{Fig7}}
\end{figure}

Time evolution of the distribution in the phase space for the TS instability with $\nu=0.001$ are shown in Fig.\ \ref{Fig7}. At time $t=10$ (Fig.\ \ref{Fig7} (a) and (d)), the initial two-stream is dominant and the center beam caused by collision start to grow. In this case, the wave energy saturates around time $t=150$. When it comes to $t=220$, the peaks of two-stream shrinks and the collision caused center beam become obvious(Fig.\ \ref{Fig7} (e)). Two vortices (Fig.\ \ref{Fig7} (b)) are formed in the phase space by the interactions among the collision caused beam, $v_1=-6$ and $v_2=6$ beams. With the collision, the collision caused beam grows and the initial two-stream shinks continously. At time $t=400$, the peak of the collision caused beam is comparable with that of the initial two-stream(Fig.\ \ref{Fig7} (f)). The two vortices sturcture in the phase space is still exsist and the center beam, which seperate the two vortices, becomes more obvious. Although the main structure in phase space is roughly clear and visible, the detail of the figure becomes more and more vague by the collision.

\begin{figure}[htbp]%[p!]%
\includegraphics[width=15cm,height=7.5cm]{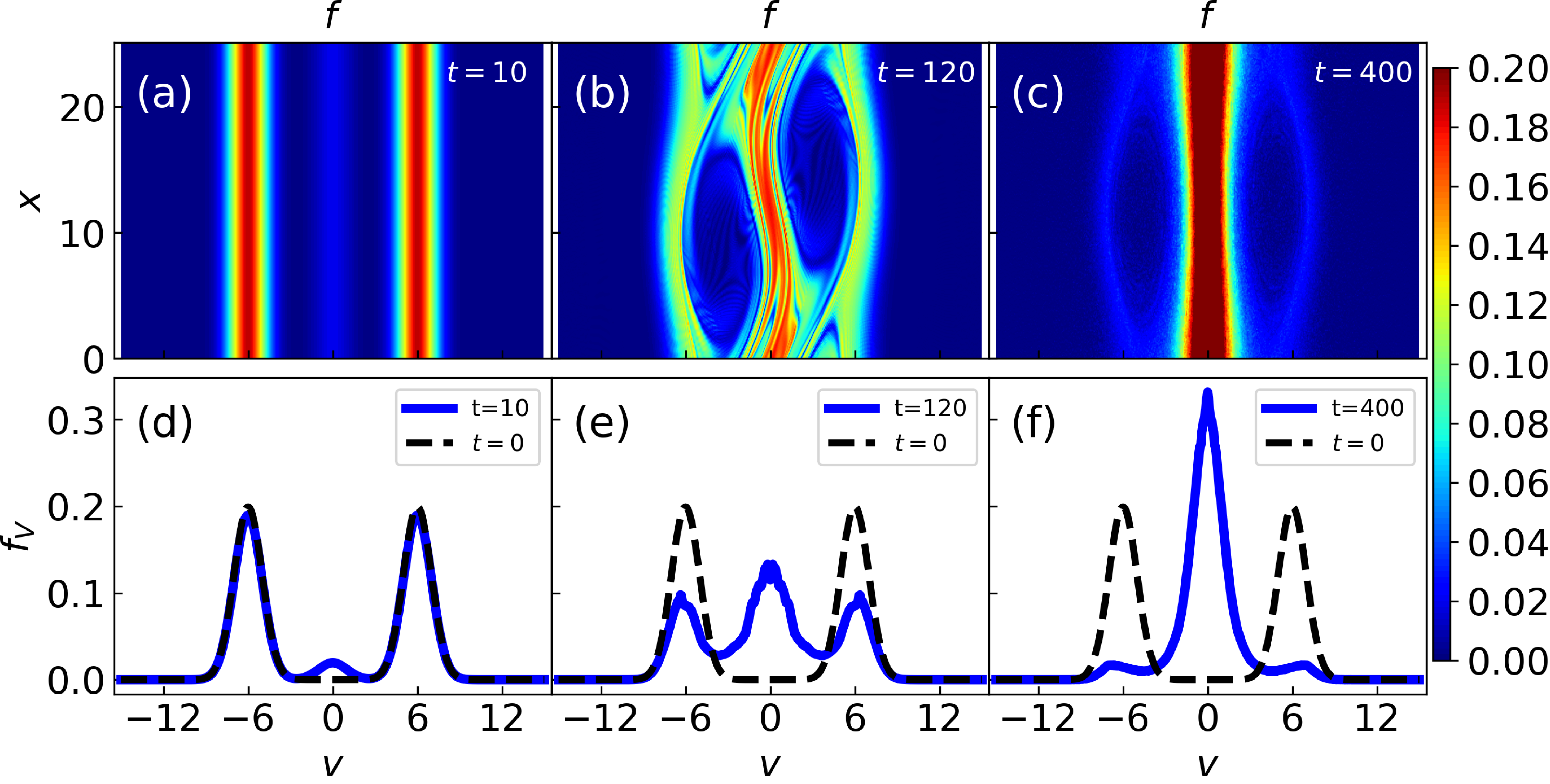}
\caption{The electron phase space (upper row) and $f_{\rm v}$ (lower row) of the two-stream instability for $\nu=0.005$ at $t=10$ [(a) and (d)], $120$ [(b) and (e)], and $400$ [(c) and (f)]. Here $f_{\rm avg}(v)$ eventually becomes single peaked as a result of collisional thermalization of the electron distribution.
\label{Fig8}}
\end{figure}

Time evolution of the distribution in the phase space for the TS instability with $\nu=0.005$ are shown in Fig.\ \ref{Fig8}. At time $t=10$ (Fig.\ \ref{Fig8} (a) and (d)), the initial two-stream is dominant and the center beam caused by collision becomes more obvious than the $\nu=0.001$ case. The wave energy saturates around time $t=100$. For $t=120$, the peaks of two-stream shrinks and the collision caused center beam become larger than the two-stream (Fig.\ \ref{Fig8} (e)). At time $t=400$, the peak of the collision caused beam is dorminant and the initial two-stream almost disappears and only little remains left (Fig.\ \ref{Fig8} (f)). The two vortices structure in the phase space caused by the collision becomes obvious (Fig.\ \ref{Fig8} (b)) and then eliminate (Fig.\ \ref{Fig8} (c)) with time. Since the collision is stronger, phase space figure becomes vague more quickly than that of the case $\nu=0.001$.

\section{Suppression of collision on the bump-on-tail instability}
\begin{figure}%[htbp]%[p!]%
\includegraphics[height=5cm,width=6cm]{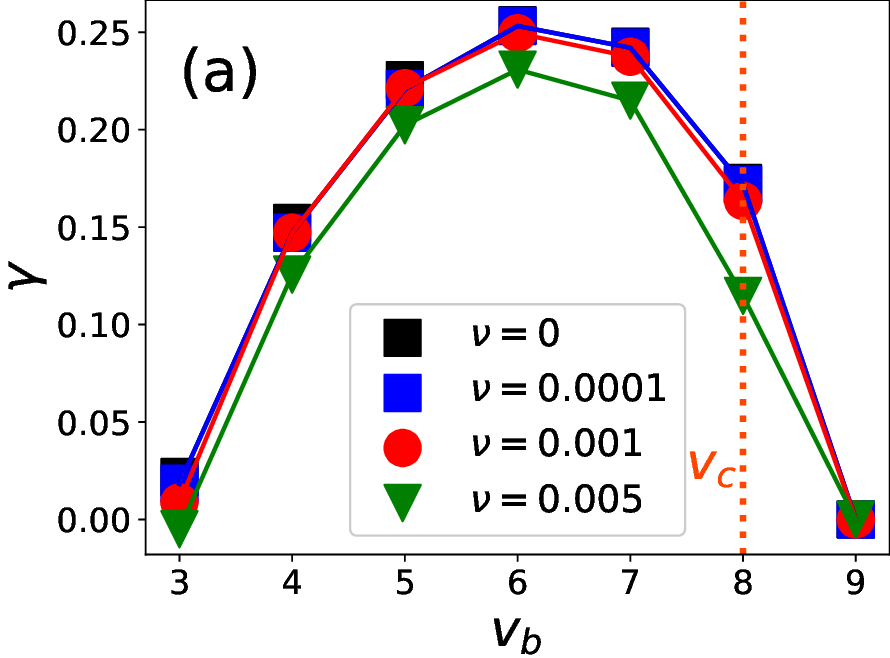}
\includegraphics[height=5cm,width=6cm]{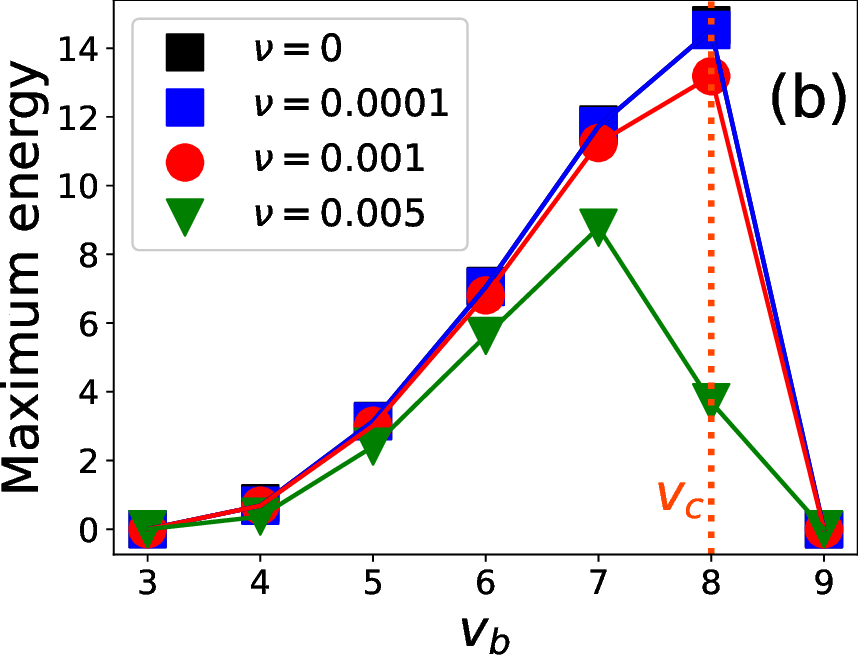}
\caption{The growth rate (a) and maximum energy (b) of the bump-on-tail instability depend on on beam velocity $v_b$. 
\label{Fig9}}
\end{figure}

Without collision, the growth rate of BOT instability increases first and then decreases with the beam velocity (Fig.\ \ref{Fig9}). 
The threshold beam velocity of the instability is $9$, which is the critical velocity plus the beam width in phase space. 
If the beam velocity is bigger than this threshold, no instability could be excited.
The maximum wave energy increases with the beam velocity and directly approaches to $0$ 
for beam velocity bigger than the critical velocity plus the beam width. 
The collision can suppress the BOT instability for the whole beam velocity region, which means both the growth rate and maximum wave energy decreases with the collision rate. When the beam velocity is set to be the critical velocity, the maximum wave energy decreases dramatically with the collision rate.

\section{Summary}
Based on our simulations of the TS and BOT instabilities using Vlasov-Poisson code with Krook collision, we find that the collision can both suppress and excite the TS instability for different beam velocity, however it can only suppress the BOT instability. The collision always dissipate the system energy, so that it can suppress the instability. When the collision induced "beam" or "bump" in the distribution function meet the instability criterion, this collision induced beam-plasma instability can grow, which brings the excitation of the instability. The competetion of suppression and excitation effect can lead to the coplex behavior of the instability for different beam velocity and collision. In this paper, we have disscussed the mechanism and given out the transition condition of the suppression and excitation. This study is 1D Vlasov-Possion simulation for electrostatic plasma, more studies should be taken to investigate the mechanism for higher dimensions and eletromagnetic plasmas. The collision excitation of instability under certain conditions can help people to understand the instabilities related phenomena in space and fusion plasmas, especially the charged partice acceleration and electromagnetic wave excitation in dense or low-temperature plasmas.

\begin{acknowledgments}
This work was supported by the Educational Department Funds for Distinguished Young Scholars of Anhui Provinces (No. 2023AH020023),
the National Natural Science Foundation of China (Nos. 12175272, 11875253 and 11205194),
the Introduced Talents and Doctors Startup Funds of Anhui Jianzhu University (No. 2022QDZ21 and 2020QDZ24),
the Fundamental Research Funds for the Central Universities (No. WK3420000004),
the Collaborative Innovation Program of Hefei Science Center, CAS (No. 2019HSC-CIP015),
the Open Project Funds for Hubei Key Laboratory of Optical Information and Pattern Recognition, Wuhan Institute of Technology (NO. 202204).
We have made use of the computing facilities of
the Supercomputing Center of the University of Science and Technology of China.
\end{acknowledgments}

\bibliographystyle{apsrev4-2}
\bibliography{Collision_Supression_Excitation_ref}

\begin{thebibliography}{40}%
\makeatletter
\providecommand \@ifxundefined [1]{%
 \@ifx{#1\undefined}
}%
\providecommand \@ifnum [1]{%
 \ifnum #1\expandafter \@firstoftwo
 \else \expandafter \@secondoftwo
 \fi
}%
\providecommand \@ifx [1]{%
 \ifx #1\expandafter \@firstoftwo
 \else \expandafter \@secondoftwo
 \fi
}%
\providecommand \natexlab [1]{#1}%
\providecommand \enquote  [1]{``#1''}%
\providecommand \bibnamefont  [1]{#1}%
\providecommand \bibfnamefont [1]{#1}%
\providecommand \citenamefont [1]{#1}%
\providecommand \href@noop [0]{\@secondoftwo}%
\providecommand \href [0]{\begingroup \@sanitize@url \@href}%
\providecommand \@href[1]{\@@startlink{#1}\@@href}%
\providecommand \@@href[1]{\endgroup#1\@@endlink}%
\providecommand \@sanitize@url [0]{\catcode `\\12\catcode `\$12\catcode
  `\&12\catcode `\#12\catcode `\^12\catcode `\_12\catcode `\%12\relax}%
\providecommand \@@startlink[1]{}%
\providecommand \@@endlink[0]{}%
\providecommand \url  [0]{\begingroup\@sanitize@url \@url }%
\providecommand \@url [1]{\endgroup\@href {#1}{\urlprefix }}%
\providecommand \urlprefix  [0]{URL }%
\providecommand \Eprint [0]{\href }%
\providecommand \doibase [0]{https://doi.org/}%
\providecommand \selectlanguage [0]{\@gobble}%
\providecommand \bibinfo  [0]{\@secondoftwo}%
\providecommand \bibfield  [0]{\@secondoftwo}%
\providecommand \translation [1]{[#1]}%
\providecommand \BibitemOpen [0]{}%
\providecommand \bibitemStop [0]{}%
\providecommand \bibitemNoStop [0]{.\EOS\space}%
\providecommand \EOS [0]{\spacefactor3000\relax}%
\providecommand \BibitemShut  [1]{\csname bibitem#1\endcsname}%
\let\auto@bib@innerbib\@empty
%</preamble>
\bibitem [{\citenamefont {Deneef}\ \emph {et~al.}(1973)\citenamefont {Deneef},
  \citenamefont {Malmberg},\ and\ \citenamefont {O'Neil}}]{Deneef1973}%
  \BibitemOpen
  \bibfield  {author} {\bibinfo {author} {\bibfnamefont {C.~P.}\ \bibnamefont
  {Deneef}}, \bibinfo {author} {\bibfnamefont {J.~H.}\ \bibnamefont
  {Malmberg}},\ and\ \bibinfo {author} {\bibfnamefont {T.~M.}\ \bibnamefont
  {O'Neil}},\ }\href@noop {} {\bibfield  {journal} {\bibinfo  {journal} {Phys.
  Rev. Lett.}\ }\textbf {\bibinfo {volume} {30}},\ \bibinfo {pages} {1032}
  (\bibinfo {year} {1973})}\BibitemShut {NoStop}%
\bibitem [{\citenamefont {Whelan}\ and\ \citenamefont
  {Stenzel}(1983)}]{Whelan1983}%
  \BibitemOpen
  \bibfield  {author} {\bibinfo {author} {\bibfnamefont {D.~A.}\ \bibnamefont
  {Whelan}}\ and\ \bibinfo {author} {\bibfnamefont {R.~L.}\ \bibnamefont
  {Stenzel}},\ }\href@noop {} {\bibfield  {journal} {\bibinfo  {journal} {Phys.
  Rev. Lett.}\ }\textbf {\bibinfo {volume} {50}},\ \bibinfo {pages} {1133}
  (\bibinfo {year} {1983})}\BibitemShut {NoStop}%
\bibitem [{\citenamefont {Hartmann}\ \emph {et~al.}(1995)\citenamefont
  {Hartmann}, \citenamefont {Driscoll}, \citenamefont {O'Neil},\ and\
  \citenamefont {Shapiro}}]{Hartmann1995}%
  \BibitemOpen
  \bibfield  {author} {\bibinfo {author} {\bibfnamefont {D.~A.}\ \bibnamefont
  {Hartmann}}, \bibinfo {author} {\bibfnamefont {C.~F.}\ \bibnamefont
  {Driscoll}}, \bibinfo {author} {\bibfnamefont {T.~M.}\ \bibnamefont
  {O'Neil}},\ and\ \bibinfo {author} {\bibfnamefont {V.~D.}\ \bibnamefont
  {Shapiro}},\ }\href@noop {} {\bibfield  {journal} {\bibinfo  {journal} {Phys.
  Plasmas}\ }\textbf {\bibinfo {volume} {2}},\ \bibinfo {pages} {654} (\bibinfo
  {year} {1995})}\BibitemShut {NoStop}%
\bibitem [{\citenamefont {Penrose}(1960)}]{Penrose1960}%
  \BibitemOpen
  \bibfield  {author} {\bibinfo {author} {\bibfnamefont {O.}~\bibnamefont
  {Penrose}},\ }\href@noop {} {\bibfield  {journal} {\bibinfo  {journal} {Phys.
  Fluids}\ }\textbf {\bibinfo {volume} {3}},\ \bibinfo {pages} {258} (\bibinfo
  {year} {1960})}\BibitemShut {NoStop}%
\bibitem [{\citenamefont {Hasegawa}(1968)}]{Hasegawa1968}%
  \BibitemOpen
  \bibfield  {author} {\bibinfo {author} {\bibfnamefont {A.}~\bibnamefont
  {Hasegawa}},\ }\href@noop {} {\bibfield  {journal} {\bibinfo  {journal}
  {Phys. Rev.}\ }\textbf {\bibinfo {volume} {169}},\ \bibinfo {pages} {204}
  (\bibinfo {year} {1968})}\BibitemShut {NoStop}%
\bibitem [{\citenamefont {Clemmow}\ and\ \citenamefont
  {Dougherty}(1969)}]{Clemmow1969}%
  \BibitemOpen
  \bibfield  {author} {\bibinfo {author} {\bibfnamefont {P.~C.}\ \bibnamefont
  {Clemmow}}\ and\ \bibinfo {author} {\bibfnamefont {J.~P.}\ \bibnamefont
  {Dougherty}},\ }\href@noop {} {\emph {\bibinfo {title} {Electrodynamics of
  Particles and Plasmas}}}\ (\bibinfo  {publisher} {Addison-Wesley Reading,
  Mass},\ \bibinfo {year} {1969})\BibitemShut {NoStop}%
\bibitem [{\citenamefont {Infeld}\ and\ \citenamefont
  {Skorupski}(1970)}]{Infeld1970}%
  \BibitemOpen
  \bibfield  {author} {\bibinfo {author} {\bibfnamefont {E.}~\bibnamefont
  {Infeld}}\ and\ \bibinfo {author} {\bibfnamefont {A.}~\bibnamefont
  {Skorupski}},\ }\href {https://doi.org/10.1017/S0022377800005286} {\bibfield
  {journal} {\bibinfo  {journal} {J. Plasma Phys.}\ }\textbf {\bibinfo {volume}
  {4}},\ \bibinfo {pages} {607} (\bibinfo {year} {1970})}\BibitemShut {NoStop}%
\bibitem [{\citenamefont {Schamel}(1982)}]{Schamel1982}%
  \BibitemOpen
  \bibfield  {author} {\bibinfo {author} {\bibfnamefont {H.}~\bibnamefont
  {Schamel}},\ }\href@noop {} {\bibfield  {journal} {\bibinfo  {journal} {Phys.
  Rev. Lett.}\ }\textbf {\bibinfo {volume} {48}},\ \bibinfo {pages} {481}
  (\bibinfo {year} {1982})}\BibitemShut {NoStop}%
\bibitem [{\citenamefont {Chen}(1984)}]{Chen1984}%
  \BibitemOpen
  \bibfield  {author} {\bibinfo {author} {\bibfnamefont {F.~F.}\ \bibnamefont
  {Chen}},\ }\href@noop {} {\emph {\bibinfo {title} {Introduction to Plasma
  Physics}}}\ (\bibinfo  {publisher} {Plenum Press, New York},\ \bibinfo {year}
  {1984})\BibitemShut {NoStop}%
\bibitem [{\citenamefont {El-Labany}\ and\ \citenamefont
  {Rowlands}(1986)}]{El-Labany1986}%
  \BibitemOpen
  \bibfield  {author} {\bibinfo {author} {\bibfnamefont {S.~K.}\ \bibnamefont
  {El-Labany}}\ and\ \bibinfo {author} {\bibfnamefont {G.}~\bibnamefont
  {Rowlands}},\ }\href@noop {} {\bibfield  {journal} {\bibinfo  {journal}
  {Plasma Phys. Control. Fusion}\ }\textbf {\bibinfo {volume} {28}},\ \bibinfo
  {pages} {1549} (\bibinfo {year} {1986})}\BibitemShut {NoStop}%
\bibitem [{\citenamefont {Rostomian}(1994)}]{Rostomian1988}%
  \BibitemOpen
  \bibfield  {author} {\bibinfo {author} {\bibfnamefont {E.~V.}\ \bibnamefont
  {Rostomian}},\ }\href@noop {} {\bibfield  {journal} {\bibinfo  {journal}
  {Plasma Phys. Control. Fusion}\ }\textbf {\bibinfo {volume} {36}},\ \bibinfo
  {pages} {1737} (\bibinfo {year} {1994})}\BibitemShut {NoStop}%
\bibitem [{\citenamefont {Boyd}\ and\ \citenamefont
  {Sanderson}(2003)}]{Boyd2003}%
  \BibitemOpen
  \bibfield  {author} {\bibinfo {author} {\bibfnamefont {T.~J.~M.}\
  \bibnamefont {Boyd}}\ and\ \bibinfo {author} {\bibfnamefont {J.~J.}\
  \bibnamefont {Sanderson}},\ }\href@noop {} {\emph {\bibinfo {title} {The
  Physics of Plasmas}}}\ (\bibinfo  {publisher} {Cambridge University Press},\
  \bibinfo {year} {2003})\BibitemShut {NoStop}%
\bibitem [{\citenamefont {Ng}\ \emph {et~al.}(2004)\citenamefont {Ng},
  \citenamefont {Bhattacharjee},\ and\ \citenamefont {Skiff}}]{Ng2004}%
  \BibitemOpen
  \bibfield  {author} {\bibinfo {author} {\bibfnamefont {C.~S.}\ \bibnamefont
  {Ng}}, \bibinfo {author} {\bibfnamefont {A.}~\bibnamefont {Bhattacharjee}},\
  and\ \bibinfo {author} {\bibfnamefont {F.}~\bibnamefont {Skiff}},\ }\href
  {https://doi.org/10.1103/PhysRevLett.92.065002} {\bibfield  {journal}
  {\bibinfo  {journal} {Phys. Rev. Lett.}\ }\textbf {\bibinfo {volume} {92}},\
  \bibinfo {pages} {065002} (\bibinfo {year} {2004})}\BibitemShut {NoStop}%
\bibitem [{\citenamefont {Roberts}\ and\ \citenamefont
  {Berk}(1967)}]{Roberts1967}%
  \BibitemOpen
  \bibfield  {author} {\bibinfo {author} {\bibfnamefont {K.~V.}\ \bibnamefont
  {Roberts}}\ and\ \bibinfo {author} {\bibfnamefont {H.~L.}\ \bibnamefont
  {Berk}},\ }\href@noop {} {\bibfield  {journal} {\bibinfo  {journal} {Phys.
  Rev. Lett.}\ }\textbf {\bibinfo {volume} {19}},\ \bibinfo {pages} {297}
  (\bibinfo {year} {1967})}\BibitemShut {NoStop}%
\bibitem [{\citenamefont {Morse}\ and\ \citenamefont
  {Nielson}(1969)}]{Morse1969b}%
  \BibitemOpen
  \bibfield  {author} {\bibinfo {author} {\bibfnamefont {R.~L.}\ \bibnamefont
  {Morse}}\ and\ \bibinfo {author} {\bibfnamefont {C.~W.}\ \bibnamefont
  {Nielson}},\ }\href@noop {} {\bibfield  {journal} {\bibinfo  {journal} {Phys.
  Rev. Lett.}\ }\textbf {\bibinfo {volume} {23}},\ \bibinfo {pages} {1087}
  (\bibinfo {year} {1969})}\BibitemShut {NoStop}%
\bibitem [{\citenamefont {Gentle}\ and\ \citenamefont
  {Hohr}(1973)}]{Gentle1973a}%
  \BibitemOpen
  \bibfield  {author} {\bibinfo {author} {\bibfnamefont {K.~W.}\ \bibnamefont
  {Gentle}}\ and\ \bibinfo {author} {\bibfnamefont {J.}~\bibnamefont {Hohr}},\
  }\href@noop {} {\bibfield  {journal} {\bibinfo  {journal} {Phys. Rev. Lett.}\
  }\textbf {\bibinfo {volume} {30}},\ \bibinfo {pages} {75} (\bibinfo {year}
  {1973})}\BibitemShut {NoStop}%
\bibitem [{\citenamefont {Lacina}\ \emph {et~al.}(1976)\citenamefont {Lacina},
  \citenamefont {Krlin},\ and\ \citenamefont {Korbel}}]{Lacina1976}%
  \BibitemOpen
  \bibfield  {author} {\bibinfo {author} {\bibfnamefont {J.}~\bibnamefont
  {Lacina}}, \bibinfo {author} {\bibfnamefont {L.}~\bibnamefont {Krlin}},\ and\
  \bibinfo {author} {\bibfnamefont {S.}~\bibnamefont {Korbel}},\ }\href@noop {}
  {\bibfield  {journal} {\bibinfo  {journal} {Plasma Phys.}\ }\textbf {\bibinfo
  {volume} {18}},\ \bibinfo {pages} {471} (\bibinfo {year} {1976})}\BibitemShut
  {NoStop}%
\bibitem [{\citenamefont {Morey}\ and\ \citenamefont
  {Boswell}(1989)}]{Morey1989}%
  \BibitemOpen
  \bibfield  {author} {\bibinfo {author} {\bibfnamefont {I.~J.}\ \bibnamefont
  {Morey}}\ and\ \bibinfo {author} {\bibfnamefont {R.~W.}\ \bibnamefont
  {Boswell}},\ }\href@noop {} {\bibfield  {journal} {\bibinfo  {journal} {Phys.
  Fluids B}\ }\textbf {\bibinfo {volume} {1}},\ \bibinfo {pages} {1502}
  (\bibinfo {year} {1989})}\BibitemShut {NoStop}%
\bibitem [{\citenamefont {Zheng}\ \emph {et~al.}(2006)\citenamefont {Zheng},
  \citenamefont {Liu}, \citenamefont {Zhang}, \citenamefont {Zhu},\ and\
  \citenamefont {He}}]{Zheng2006}%
  \BibitemOpen
  \bibfield  {author} {\bibinfo {author} {\bibfnamefont {C.~Y.}\ \bibnamefont
  {Zheng}}, \bibinfo {author} {\bibfnamefont {Z.~J.}\ \bibnamefont {Liu}},
  \bibinfo {author} {\bibfnamefont {A.~Q.}\ \bibnamefont {Zhang}}, \bibinfo
  {author} {\bibfnamefont {S.~P.}\ \bibnamefont {Zhu}},\ and\ \bibinfo {author}
  {\bibfnamefont {X.~T.}\ \bibnamefont {He}},\ }\href
  {https://doi.org/10.1017/S0022377805003971} {\bibfield  {journal} {\bibinfo
  {journal} {J. Plasma Phys.}\ }\textbf {\bibinfo {volume} {72}},\ \bibinfo
  {pages} {249} (\bibinfo {year} {2006})}\BibitemShut {NoStop}%
\bibitem [{\citenamefont {Dieckmann}\ \emph {et~al.}(2006)\citenamefont
  {Dieckmann}, \citenamefont {Eliasson}, \citenamefont {Shukla}, \citenamefont
  {Sircombe},\ and\ \citenamefont {Dendy}}]{Dieckmann2006}%
  \BibitemOpen
  \bibfield  {author} {\bibinfo {author} {\bibfnamefont {M.~E.}\ \bibnamefont
  {Dieckmann}}, \bibinfo {author} {\bibfnamefont {B.}~\bibnamefont {Eliasson}},
  \bibinfo {author} {\bibfnamefont {P.~K.}\ \bibnamefont {Shukla}}, \bibinfo
  {author} {\bibfnamefont {N.~J.}\ \bibnamefont {Sircombe}},\ and\ \bibinfo
  {author} {\bibfnamefont {R.~O.}\ \bibnamefont {Dendy}},\ }\href@noop {}
  {\bibfield  {journal} {\bibinfo  {journal} {Plasma Phys. Control. Fusion.}\
  }\textbf {\bibinfo {volume} {48}},\ \bibinfo {pages} {B303} (\bibinfo {year}
  {2006})}\BibitemShut {NoStop}%
\bibitem [{\citenamefont {Daldorff}\ \emph {et~al.}(2011)\citenamefont
  {Daldorff}, \citenamefont {Pecseli}, \citenamefont {Trulsen}, \citenamefont
  {Ulriksen}, \citenamefont {Eliasson},\ and\ \citenamefont
  {Stenflo}}]{Daldorff2011}%
  \BibitemOpen
  \bibfield  {author} {\bibinfo {author} {\bibfnamefont {L.~K.~S.}\
  \bibnamefont {Daldorff}}, \bibinfo {author} {\bibfnamefont {H.~L.}\
  \bibnamefont {Pecseli}}, \bibinfo {author} {\bibfnamefont {J.~K.}\
  \bibnamefont {Trulsen}}, \bibinfo {author} {\bibfnamefont {M.~I.}\
  \bibnamefont {Ulriksen}}, \bibinfo {author} {\bibfnamefont {B.}~\bibnamefont
  {Eliasson}},\ and\ \bibinfo {author} {\bibfnamefont {L.}~\bibnamefont
  {Stenflo}},\ }\href@noop {} {\bibfield  {journal} {\bibinfo  {journal} {Phys.
  Plasmas}\ }\textbf {\bibinfo {volume} {18}},\ \bibinfo {pages} {052107}
  (\bibinfo {year} {2011})}\BibitemShut {NoStop}%
\bibitem [{\citenamefont {Volokitin}\ and\ \citenamefont
  {Krafft}(2012)}]{Volokitin2012}%
  \BibitemOpen
  \bibfield  {author} {\bibinfo {author} {\bibfnamefont {A.}~\bibnamefont
  {Volokitin}}\ and\ \bibinfo {author} {\bibfnamefont {C.}~\bibnamefont
  {Krafft}},\ }\href@noop {} {\bibfield  {journal} {\bibinfo  {journal} {Plasma
  Phys. Control. Fusion}\ }\textbf {\bibinfo {volume} {54}},\ \bibinfo {pages}
  {085002} (\bibinfo {year} {2012})}\BibitemShut {NoStop}%
\bibitem [{\citenamefont {Tsiklauri}(2011)}]{Tsiklauri2011}%
  \BibitemOpen
  \bibfield  {author} {\bibinfo {author} {\bibfnamefont {D.}~\bibnamefont
  {Tsiklauri}},\ }\href {https://doi.org/10.1063/1.3590928} {\bibfield
  {journal} {\bibinfo  {journal} {Phys. Plasmas}\ }\textbf {\bibinfo {volume}
  {18}},\ \bibinfo {pages} {052903} (\bibinfo {year} {2011})}\BibitemShut
  {NoStop}%
\bibitem [{\citenamefont {Thurgood}\ and\ \citenamefont
  {Tsiklauri}(2015)}]{Thurgood2015}%
  \BibitemOpen
  \bibfield  {author} {\bibinfo {author} {\bibfnamefont {J.~O.}\ \bibnamefont
  {Thurgood}}\ and\ \bibinfo {author} {\bibfnamefont {D.}~\bibnamefont
  {Tsiklauri}},\ }\href@noop {} {\bibfield  {journal} {\bibinfo  {journal}
  {Astronomy \& Astrophysics}\ }\textbf {\bibinfo {volume} {584}},\ \bibinfo
  {pages} {A83} (\bibinfo {year} {2015})}\BibitemShut {NoStop}%
\bibitem [{\citenamefont {Thurgood}\ and\ \citenamefont
  {Tsiklauri}(2016)}]{Thurgood2016}%
  \BibitemOpen
  \bibfield  {author} {\bibinfo {author} {\bibfnamefont {J.~O.}\ \bibnamefont
  {Thurgood}}\ and\ \bibinfo {author} {\bibfnamefont {D.}~\bibnamefont
  {Tsiklauri}},\ }\href@noop {} {\bibfield  {journal} {\bibinfo  {journal} {J.
  Plasma Phys.}\ }\textbf {\bibinfo {volume} {82}} (\bibinfo {year}
  {2016})}\BibitemShut {NoStop}%
\bibitem [{\citenamefont {Alves}\ \emph {et~al.}(2014)\citenamefont {Alves},
  \citenamefont {Grismayer}, \citenamefont {Fonseca},\ and\ \citenamefont
  {Silva}}]{Alves2014}%
  \BibitemOpen
  \bibfield  {author} {\bibinfo {author} {\bibfnamefont {E.~P.}\ \bibnamefont
  {Alves}}, \bibinfo {author} {\bibfnamefont {T.}~\bibnamefont {Grismayer}},
  \bibinfo {author} {\bibfnamefont {R.~A.}\ \bibnamefont {Fonseca}},\ and\
  \bibinfo {author} {\bibfnamefont {L.~O.}\ \bibnamefont {Silva}},\ }\href
  {https://doi.org/10.1088/1367-2630/16/3/035007} {\bibfield  {journal}
  {\bibinfo  {journal} {New J. Phys.}\ }\textbf {\bibinfo {volume} {16}},\
  \bibinfo {pages} {035007} (\bibinfo {year} {2014})}\BibitemShut {NoStop}%
\bibitem [{\citenamefont {Freethy}\ \emph {et~al.}(2015)\citenamefont
  {Freethy}, \citenamefont {McClements}, \citenamefont {Chapman}, \citenamefont
  {Dendy}, \citenamefont {Lai}, \citenamefont {Pamela}, \citenamefont
  {Shevchenko},\ and\ \citenamefont {Vann}}]{Freethy2015}%
  \BibitemOpen
  \bibfield  {author} {\bibinfo {author} {\bibfnamefont {S.~J.}\ \bibnamefont
  {Freethy}}, \bibinfo {author} {\bibfnamefont {K.~G.}\ \bibnamefont
  {McClements}}, \bibinfo {author} {\bibfnamefont {S.~C.}\ \bibnamefont
  {Chapman}}, \bibinfo {author} {\bibfnamefont {R.~O.}\ \bibnamefont {Dendy}},
  \bibinfo {author} {\bibfnamefont {W.~N.}\ \bibnamefont {Lai}}, \bibinfo
  {author} {\bibfnamefont {S.~J.~P.}\ \bibnamefont {Pamela}}, \bibinfo {author}
  {\bibfnamefont {V.~F.}\ \bibnamefont {Shevchenko}},\ and\ \bibinfo {author}
  {\bibfnamefont {R.~G.~L.}\ \bibnamefont {Vann}},\ }\href
  {https://doi.org/10.1103/PhysRevLett.114.125004} {\bibfield  {journal}
  {\bibinfo  {journal} {Phys. Rev. Lett.}\ }\textbf {\bibinfo {volume} {114}},\
  \bibinfo {pages} {125004} (\bibinfo {year} {2015})}\BibitemShut {NoStop}%
\bibitem [{\citenamefont {Kemp}\ \emph {et~al.}(2006)\citenamefont {Kemp},
  \citenamefont {Sentoku}, \citenamefont {Sotnikov},\ and\ \citenamefont
  {Wilks}}]{Kemp2006}%
  \BibitemOpen
  \bibfield  {author} {\bibinfo {author} {\bibfnamefont {A.~J.}\ \bibnamefont
  {Kemp}}, \bibinfo {author} {\bibfnamefont {Y.}~\bibnamefont {Sentoku}},
  \bibinfo {author} {\bibfnamefont {V.}~\bibnamefont {Sotnikov}},\ and\
  \bibinfo {author} {\bibfnamefont {S.~C.}\ \bibnamefont {Wilks}},\ }\href
  {https://doi.org/10.1103/PhysRevLett.97.235001} {\bibfield  {journal}
  {\bibinfo  {journal} {Phys. Rev. Lett.}\ }\textbf {\bibinfo {volume} {97}},\
  \bibinfo {pages} {235001} (\bibinfo {year} {2006})}\BibitemShut {NoStop}%
\bibitem [{\citenamefont {Hao}\ \emph {et~al.}(2009)\citenamefont {Hao},
  \citenamefont {Sheng}, \citenamefont {Ren},\ and\ \citenamefont
  {Zhang}}]{Hao2009}%
  \BibitemOpen
  \bibfield  {author} {\bibinfo {author} {\bibfnamefont {B.}~\bibnamefont
  {Hao}}, \bibinfo {author} {\bibfnamefont {Z.-M.}\ \bibnamefont {Sheng}},
  \bibinfo {author} {\bibfnamefont {C.}~\bibnamefont {Ren}},\ and\ \bibinfo
  {author} {\bibfnamefont {J.}~\bibnamefont {Zhang}},\ }\href
  {https://doi.org/10.1103/PhysRevE.79.046409} {\bibfield  {journal} {\bibinfo
  {journal} {Phys. Rev. E}\ }\textbf {\bibinfo {volume} {79}},\ \bibinfo
  {pages} {046409} (\bibinfo {year} {2009})}\BibitemShut {NoStop}%
\bibitem [{\citenamefont {Zhou}\ and\ \citenamefont {Bellan}(2023)}]{Zhou2023}%
  \BibitemOpen
  \bibfield  {author} {\bibinfo {author} {\bibfnamefont {Y.}~\bibnamefont
  {Zhou}}\ and\ \bibinfo {author} {\bibfnamefont {P.~M.}\ \bibnamefont
  {Bellan}},\ }\href {https://doi.org/10.1063/5.0146806} {\bibfield  {journal}
  {\bibinfo  {journal} {Phys. Plasmas}\ }\textbf {\bibinfo {volume} {30}},\
  \bibinfo {pages} {052101} (\bibinfo {year} {2023})}\BibitemShut {NoStop}%
\bibitem [{\citenamefont {Musha}\ and\ \citenamefont
  {Yoshida}(1964)}]{Musha1964}%
  \BibitemOpen
  \bibfield  {author} {\bibinfo {author} {\bibfnamefont {T.}~\bibnamefont
  {Musha}}\ and\ \bibinfo {author} {\bibfnamefont {F.}~\bibnamefont
  {Yoshida}},\ }\href@noop {} {\bibfield  {journal} {\bibinfo  {journal} {Phys.
  Rev.}\ }\textbf {\bibinfo {volume} {133}},\ \bibinfo {pages} {1303} (\bibinfo
  {year} {1964})}\BibitemShut {NoStop}%
\bibitem [{\citenamefont {Stenflo}(1968)}]{Stenflo1968}%
  \BibitemOpen
  \bibfield  {author} {\bibinfo {author} {\bibfnamefont {L.}~\bibnamefont
  {Stenflo}},\ }\href {https://doi.org/10.1088/0032-1028/10/8/407} {\bibfield
  {journal} {\bibinfo  {journal} {Plasma Phys.}\ }\textbf {\bibinfo {volume}
  {10}},\ \bibinfo {pages} {801} (\bibinfo {year} {1968})}\BibitemShut
  {NoStop}%
\bibitem [{\citenamefont {Hou}\ \emph {et~al.}(2015{\natexlab{a}})\citenamefont
  {Hou}, \citenamefont {Chen}, \citenamefont {Yu},\ and\ \citenamefont
  {Wu}}]{Hou2015b}%
  \BibitemOpen
  \bibfield  {author} {\bibinfo {author} {\bibfnamefont {Y.~W.}\ \bibnamefont
  {Hou}}, \bibinfo {author} {\bibfnamefont {M.~X.}\ \bibnamefont {Chen}},
  \bibinfo {author} {\bibfnamefont {M.~Y.}\ \bibnamefont {Yu}},\ and\ \bibinfo
  {author} {\bibfnamefont {B.}~\bibnamefont {Wu}},\ }\href@noop {} {\bibfield
  {journal} {\bibinfo  {journal} {J. Plasma Phys.}\ }\textbf {\bibinfo {volume}
  {81}},\ \bibinfo {pages} {905810602} (\bibinfo {year}
  {2015}{\natexlab{a}})}\BibitemShut {NoStop}%
\bibitem [{\citenamefont {Hou}\ \emph {et~al.}(2024)\citenamefont {Hou},
  \citenamefont {Yu}, \citenamefont {Wang}, \citenamefont {Liu}, \citenamefont
  {Chen},\ and\ \citenamefont {Wu}}]{Hou2024}%
  \BibitemOpen
  \bibfield  {author} {\bibinfo {author} {\bibfnamefont {Y.~W.}\ \bibnamefont
  {Hou}}, \bibinfo {author} {\bibfnamefont {M.~Y.}\ \bibnamefont {Yu}},
  \bibinfo {author} {\bibfnamefont {J.~F.}\ \bibnamefont {Wang}}, \bibinfo
  {author} {\bibfnamefont {C.~Y.}\ \bibnamefont {Liu}}, \bibinfo {author}
  {\bibfnamefont {M.~X.}\ \bibnamefont {Chen}},\ and\ \bibinfo {author}
  {\bibfnamefont {B.}~\bibnamefont {Wu}},\ }\href
  {https://doi.org/10.1063/5.0181606} {\bibfield  {journal} {\bibinfo
  {journal} {Physics of Plasmas}\ }\textbf {\bibinfo {volume} {31}},\ \bibinfo
  {pages} {042102} (\bibinfo {year} {2024})}\BibitemShut {NoStop}%
\bibitem [{\citenamefont {Chen}(2006)}]{Chen2006}%
  \BibitemOpen
  \bibfield  {author} {\bibinfo {author} {\bibfnamefont {F.~F.}\ \bibnamefont
  {Chen}},\ }\href@noop {} {\emph {\bibinfo {title} {{I}ntroduction to {P}lasma
  {P}hysics and {C}ontrolled {F}usion, 2nd {E}dition}}}\ (\bibinfo  {publisher}
  {Springer, New York},\ \bibinfo {year} {2006})\BibitemShut {NoStop}%
\bibitem [{\citenamefont {Cheng}\ and\ \citenamefont
  {Knorr}(1976)}]{Cheng1976}%
  \BibitemOpen
  \bibfield  {author} {\bibinfo {author} {\bibfnamefont {C.~Z.}\ \bibnamefont
  {Cheng}}\ and\ \bibinfo {author} {\bibfnamefont {G.}~\bibnamefont {Knorr}},\
  }\href@noop {} {\bibfield  {journal} {\bibinfo  {journal} {J. Comput. Phys.}\
  }\textbf {\bibinfo {volume} {22}},\ \bibinfo {pages} {330} (\bibinfo {year}
  {1976})}\BibitemShut {NoStop}%
\bibitem [{\citenamefont {Hou}\ \emph {et~al.}(2011{\natexlab{a}})\citenamefont
  {Hou}, \citenamefont {Ma},\ and\ \citenamefont {Yu}}]{Hou2011a}%
  \BibitemOpen
  \bibfield  {author} {\bibinfo {author} {\bibfnamefont {Y.~W.}\ \bibnamefont
  {Hou}}, \bibinfo {author} {\bibfnamefont {Z.~W.}\ \bibnamefont {Ma}},\ and\
  \bibinfo {author} {\bibfnamefont {M.~Y.}\ \bibnamefont {Yu}},\ }\href@noop {}
  {\bibfield  {journal} {\bibinfo  {journal} {Phys. Plasmas}\ }\textbf
  {\bibinfo {volume} {18}},\ \bibinfo {pages} {082101} (\bibinfo {year}
  {2011}{\natexlab{a}})}\BibitemShut {NoStop}%
\bibitem [{\citenamefont {Hou}\ \emph {et~al.}(2011{\natexlab{b}})\citenamefont
  {Hou}, \citenamefont {Ma},\ and\ \citenamefont {Yu}}]{Hou2011b}%
  \BibitemOpen
  \bibfield  {author} {\bibinfo {author} {\bibfnamefont {Y.~W.}\ \bibnamefont
  {Hou}}, \bibinfo {author} {\bibfnamefont {Z.~W.}\ \bibnamefont {Ma}},\ and\
  \bibinfo {author} {\bibfnamefont {M.~Y.}\ \bibnamefont {Yu}},\ }\href@noop {}
  {\bibfield  {journal} {\bibinfo  {journal} {Phys. Plasmas}\ }\textbf
  {\bibinfo {volume} {18}},\ \bibinfo {pages} {012128} (\bibinfo {year}
  {2011}{\natexlab{b}})}\BibitemShut {NoStop}%
\bibitem [{\citenamefont {Hou}\ \emph {et~al.}(2015{\natexlab{b}})\citenamefont
  {Hou}, \citenamefont {Chen}, \citenamefont {Yu}, \citenamefont {Wu},\ and\
  \citenamefont {Wu}}]{Hou2015a}%
  \BibitemOpen
  \bibfield  {author} {\bibinfo {author} {\bibfnamefont {Y.~W.}\ \bibnamefont
  {Hou}}, \bibinfo {author} {\bibfnamefont {M.~X.}\ \bibnamefont {Chen}},
  \bibinfo {author} {\bibfnamefont {M.~Y.}\ \bibnamefont {Yu}}, \bibinfo
  {author} {\bibfnamefont {B.}~\bibnamefont {Wu}},\ and\ \bibinfo {author}
  {\bibfnamefont {Y.~C.}\ \bibnamefont {Wu}},\ }\href@noop {} {\bibfield
  {journal} {\bibinfo  {journal} {Phys. Plasmas}\ }\textbf {\bibinfo {volume}
  {22}},\ \bibinfo {pages} {45} (\bibinfo {year}
  {2015}{\natexlab{b}})}\BibitemShut {NoStop}%
\bibitem [{\citenamefont {Hou}\ \emph {et~al.}(2016)\citenamefont {Hou},
  \citenamefont {Chen}, \citenamefont {Yu},\ and\ \citenamefont
  {Wu}}]{Hou2016}%
  \BibitemOpen
  \bibfield  {author} {\bibinfo {author} {\bibfnamefont {Y.~W.}\ \bibnamefont
  {Hou}}, \bibinfo {author} {\bibfnamefont {M.~X.}\ \bibnamefont {Chen}},
  \bibinfo {author} {\bibfnamefont {M.~Y.}\ \bibnamefont {Yu}},\ and\ \bibinfo
  {author} {\bibfnamefont {B.}~\bibnamefont {Wu}},\ }\href@noop {} {\bibfield
  {journal} {\bibinfo  {journal} {Plasma Phys. Rep.}\ }\textbf {\bibinfo
  {volume} {42}},\ \bibinfo {pages} {900} (\bibinfo {year} {2016})}\BibitemShut
  {NoStop}%
\end{thebibliography}%

%\end{CJK*}
\end{document}